\documentclass[%
 reprint,
superscriptaddress,
 amsmath,amssymb,
 aps,physrev,
floatfix,
]{revtex4-2}
\pdfoutput=1
\usepackage{diagbox}
\usepackage{multirow}
\usepackage[dvipsnames]{xcolor}
\usepackage{ bbold }

\usepackage{braket}
\usepackage{graphicx}% Include figure files
\usepackage{dcolumn}% Align table columns on the decimal point
\usepackage{bm}% bold math
\usepackage{comment}
\usepackage{xcolor}

\newcommand{\hsigma}{\hat{\sigma}}
\begin{document}

\preprint{APS/123-QED}

\title{Native Three-Body Interactions\\
in a Superconducting Lattice Gauge Quantum Simulator}

%\title{Realizing Three-Qubit Interactions for Lattice Gauge Theory Simulations}% Force line breaks with \\

\author{J.H. Busnaina}
\affiliation{Institute for Quantum Computing and Department of Electrical \& Computer Engineering, University of Waterloo, Waterloo, Ontario, N2L 3G1, Canada}
\author{Z. Shi}
\affiliation{Institute for Quantum Computing and Department of Electrical \& Computer Engineering, University of Waterloo, Waterloo, Ontario, N2L 3G1, Canada}
\author{Jesús M. Alcaine-Cuervo}
\affiliation{EHU Quantum Center and Department of Physical Chemistry, University of the Basque Country UPV/EHU, P.O. Box 644, 48080 Bilbao, Spain}
\affiliation{BCMaterials, Basque Center for Materials, Applications and Nanostructures, UPV/EHU Science Park, 48940 Leioa, Spain}
\author{Cindy X.C. Yang}
\affiliation{Institute for Quantum Computing and Department of Electrical \& Computer Engineering, University of Waterloo, Waterloo, Ontario, N2L 3G1, Canada}
\author{I. Nsanzineza}
\affiliation{Institute for Quantum Computing and Department of Electrical \& Computer Engineering, University of Waterloo, Waterloo, Ontario, N2L 3G1, Canada}
\author{E. Rico}
\affiliation{EHU Quantum Center and Department of Physical Chemistry, University of the Basque Country UPV/EHU, P.O. Box 644, 48080 Bilbao, Spain}
\affiliation{DIPC - Donostia International Physics Center, Paseo Manuel de Lardizabal 4, 20018 San Sebastián, Spain}
\affiliation{European Organization for Nuclear Research (CERN), Geneva 1211, Switzerland}
\affiliation{IKERBASQUE, Basque Foundation for Science, Plaza Euskadi 5, 48009 Bilbao, Spain}
\author{C.M. Wilson}
\email{chris.wilson@uwaterloo.ca}
\affiliation{Institute for Quantum Computing and Department of Electrical \& Computer Engineering, University of Waterloo, Waterloo, Ontario, N2L 3G1, Canada}

\date{\today}% It is always \today, today,
             %  but any date may be explicitly specified

\begin{abstract}

While universal quantum computers remain under development, analog quantum simulators offer a powerful alternative for understanding complex systems in condensed matter, chemistry, and high-energy physics. One compelling application is the characterization of real-time lattice gauge theories (LGTs). LGTs are nonperturbative tools, utilizing discretized spacetime to describe gauge-invariant models. They hold immense potential for understanding fundamental physics but require enforcing local constraints analogous to electromagnetism's Gauss's Law. These constraints, which arise from gauge symmetries and dictate the form of the interaction between matter and gauge fields, are a significant challenge for simulators to enforce. Implementing these constraints at the hardware level in analog simulations is crucial. This requires realizing multibody interactions between matter and gauge-field elements, enabling them to evolve together while suppressing unwanted two-body interactions that violate the gauge symmetry. In this paper, we propose and implement a novel parametrically activated three-qubit interaction within a circuit quantum electrodynamics architecture. We experimentally demonstrate a minimal $U(1)$ spin-1/2 model with a time evolution that intrinsically satisfies Gauss's law in the system. This design serves as the foundational block for simulating LGTs on a superconducting photonic lattice.

\end{abstract}

%\keywords{Suggested keywords}%Use showkeys class option if keyword
                              %display desired
\maketitle

%\tableofcontents

\section{\label{sec:intro} Introduction}
Lattice gauge theories (LGTs) were originally conceived to tackle the strong interaction regime in quantum chromodynamics (QCD) where traditional perturbative methods break down~\cite{wilson1974quarks}.  Formulated on a discretized spacetime lattice, they are a type of field theory characterized by local (gauge) symmetries, which dictate how the theory behaves under specific transformations. These symmetries underpin the description of interactions between matter particles, mediated by dynamical force fields known as gauge fields. Over the past decades, the applications of LGTs have expanded beyond the realm of elementary particle physics into condensed matter physics~\cite{kogut1979,Wiese2013LGT,Fradkin2013field}. In this domain, they are used to describe the low-energy effective theories of various strongly correlated systems, where emergent gauge fields play a crucial role~\cite{Balents2010spinliquids,Savary2016QSL,Wen2004QFT,Prosko2017simpleZ2}.

Quantum Monte Carlo simulations have become a cornerstone for studying LGTs because of their ability to tackle nonperturbative aspects of QCD directly. These simulations have yielded invaluable insights into various aspects of QCD, including low-energy spectra~\cite{fodor12, usqcd19}, phase diagrams~\cite{detar09,fukushima10, philipsen19}, contributions to the muon magnetic moment~\cite{borsanyi21}, and plaquette operators~\cite{creutz83}. Despite their remarkable efficiency, Monte Carlo methods still face limitations in simulating crucial LGT problems~\cite{gattringer09, calzetta09}. These limitations are primarily due to the notorious ``sign problem'', which impedes convergence and even the proper definition of the simulation in specific contexts~\cite{dasgupta22,SZ2LGT18,deforcrand10}.

Quantum simulation has emerged as a promising tool for investigating LGTs in nonperturbative regimes where quantum Monte Carlo simulations are not efficient~\cite{Banuls20,Davoudi20,Klco20, Davoudi21}. This approach simulates the LGT's Hamiltonian digitally or by mimicking it with engineered quantum systems. However, digitally simulating even simple LGTs, like the one-dimensional $\mathbb{Z}_2$ model, requires breaking down the interaction term into numerous two-qubit and single-qubit gates~\cite{probing}. This complexity rapidly escalates when the local Hilbert space dimension of the gauge field is greater than two, even with hardware-efficient schemes. Additionally, encoding the vast bosonic Hilbert space of gauge fields using qubits is inefficient and requires considerable overhead. While trapped-ion systems have simulated LGTs by analytically integrating out the gauge fields~\cite{Martinez16, Kokail19, DQSofSM}, this method is limited to one dimension and requires additional complex long-range qubit interactions, which are particularly hard to realize in superconducting circuits. Therefore, analog quantum simulation (AQS) has gained significant traction for studying LGTs~\cite{Bauer23,dasgupta22, Ciavarella23, Wilkinson20, Jensen22, Homeier23, hung21, busnaina24}.

%Quantum link models (QLMs)~\cite{marcos13} offer a convenient framework for simulating LGTs using a finite-dimensional Hilbert space for the gauge fields while keeping the continuous local symmetries. These models generalize LGTs and find applications in condensed matter physics and quantum information theory~\cite{Wiese2013LGT}. In QLMs, gauge fields are replaced with spin-$S$ degrees of freedom, effectively truncating the infinite-dimensional Hilbert space of the gauge field to a finite $2S+1$ dimension~\cite{Banerjee12}.

In previous work, we have demonstrated a flexible platform for analog quantum simulation using microwave photonic lattices based on superconducting circuits~\cite{Simoen15, Sandbo18, Alaeian19, sandbo3photon, hung21, busnaina24}. In this work, we extend the platform, enabling the simulation of LGTs using superconducting quantum circuits. Our approach realizes the dynamical matter-gauge interaction as a three-qubit interaction on a photonic lattice that represents both the matter and gauge degrees of freedom. Within our basic building block, two qubits represent the matter field and the third represents the dynamical gauge field. In this finite-dimensional formulation of the gauge field, the continuous $U(1)$ gauge symmetry is preserved, and it yields simple yet physically rich LGT models that exhibit properties such as confinement-deconfinement phase transitions and spontaneous symmetry breaking~\cite{DLZ2,probing}. We demonstrate the capability of the platform by observing Gauss's law in the $U(1)$ spin-1/2 model~\cite{Celi20}, which describes emergent gauge fields relevant to quantum spin liquids~\cite{Hermele04}. Importantly, replacing the gauge qubit with an oscillator can easily extend our basic building block design to models with larger gauge Hilbert spaces~\cite{bosonic_qiskit,liu24,crane24}. Further, it allows for straightforward assembly into larger arrays, enabling the simulation of more complex LGTs~\cite{Banerjee13,marcos13}.

\section{\label{sec:th} Three-Qubit interaction for Dynamical Gauge Fields }
\begin{figure}
\centering
\includegraphics[width=\linewidth]{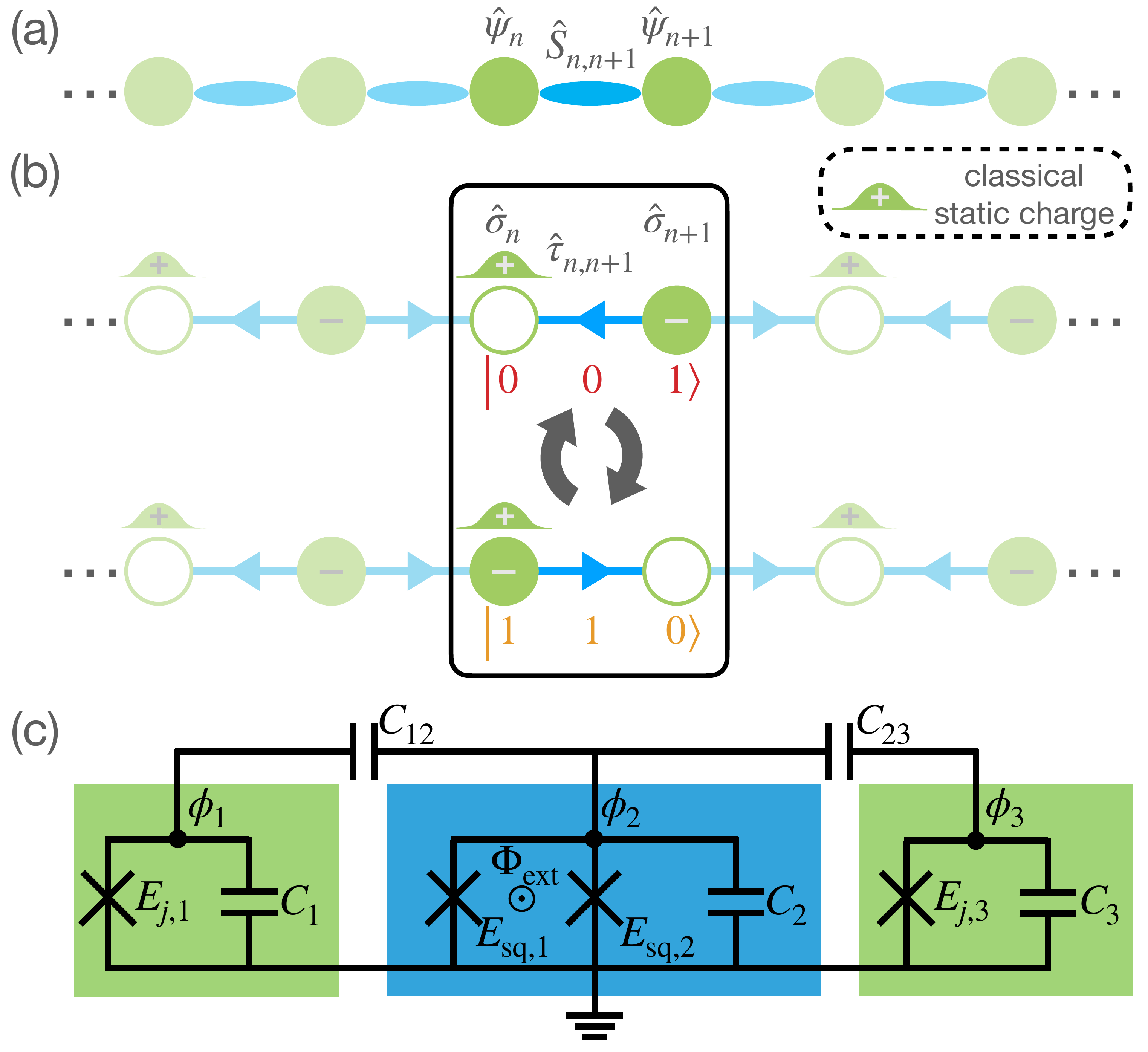}
% Previous version in main_Nov02
\caption{Building block for the analog quantum simulation of LGTs using superconducting quantum circuits. (a) Quantum link model (QLM) of one-dimensional LGTs. Matter fields, $\hat{\psi}_n$, live on lattice sites (green), and gauge fields, $\hat{S}_{n,n+1}$, live on the links (blue) connecting the sites. In QLMs, bosonic $U(1)$ gauge fields are compactly represented by spin-$S$ systems. (b) Spin representation of a $U(1)$ spin-$1/2$ QLM in one dimension, and the two spin configurations that satisfy Gauss's law in a system with two matter sites (center box). The Jordan-Wigner transformation maps the matter fields to $S=1/2$ spins. The absence (presence) of a negatively charged fermion on a matter site is indicated by an empty (filled) circle and is encoded as the $\ket{0}$ ($\ket{1}$) state of the corresponding matter qubit. A static positive charge background exists on odd matter sites. A gauge field pointing to the left (right) is encoded as the $\ket{0}$ ($\ket{1}$) state of the corresponding gauge qubit. The three-qubit terms in Eq.~\eqref{eq:H} allow a charge to hop from one matter site to its neighbor while flipping the gauge field on the link. Following this representation, in a system with two matter sites, the only two states satisfying Gauss's law in Eq.~\eqref{eq:Gsymm} are $\ket{001}$ and $\ket{110}$; they are coupled to each other by the three-qubit interaction. (c) Circuit schematic of the three-qubit device. Qubits 1 and 3 are fixed-frequency transmon qubits, i.e., a Josephson junction shunted by a capacitor, representing the matter sites. Qubit 2 is a tunable transmon qubit, i.e., a SQUID shunted by a capacitor. The qubits are coupled with capacitors $C_{12}$ and  $C_{23}$. The external flux, $\hat{\Phi}_{\mathrm{ext}}$, is applied to activate the three-body interaction parametrically.
}
\label{fig:LGTs}
\end{figure}
%(b) Depiction of simplified LGTs in Eq.~\ref{eq:H}. The matter fields are mapped to spin $1/2$ using the Jordan-Wigner transformation, and the gauge field is set as spin-$1/2$. Empty and filled sites indicate the absence or presence of quantum negative charges which are encoded as a qubit ground or excited states, respectively. The gauge field has two possible states: pointing to the left, which is encoded as qubit ground state, and pointing to the right, encoded as qubit excited state. The model involves staggered fermions which are here replaced by adding a classical positive charge background on the odd sublattice.
%(d) Cartoon of the two possible system configurations of 2-site $U(1)$ LGT. In the left panel, an electron occupies the right matter site, and the electric field points to the left, satisfying Gauss's Law. Following the representation in (b), this configuration is encoded in the system as the state $\ket{001}$. Alternatively, the electron can occupy the left matter site, encoded as $\ket{110}$.
In LGTs formulated using the Kogut-Susskind Hamiltonian approach~\cite{kogut1979, kogutsusskind}, spatial dimensions are discretized into a lattice. Matter fields reside on the lattice sites, while bosonic gauge fields reside on the links connecting these sites. As a convenient framework for simulating LGTs, quantum link models (QLMs) further replace gauge fields with spin-$S$ degrees of freedom (Fig.~\ref{fig:LGTs}(a)), effectively truncating the infinite-dimensional Hilbert space of the gauge field to a finite $2S+1$ dimension while keeping the continuous local symmetries~\cite{marcos13,Banerjee12}. These models generalize LGTs and find applications in condensed matter physics and quantum information theory~\cite{Wiese2013LGT}.

Here, we consider spin-$1/2$ links such that the link spins become Pauli operators $\hat{\tau}^i_{n,n+1}(i=x,y,z)$, and the gauge energy term is a constant, $(\hat{\tau}^z_{n,n+1})^2 = 1$. While superconducting circuits and many other platforms lack inherently fermionic building blocks, one can map fermions in a one-dimensional lattice to spin-$1/2$ sites using the Jordan-Wigner transformation~\cite{yang2016ion}. The final (1+1)D $U(1)$ LGT Hamiltonian, depicted in Fig.~\ref{fig:LGTs}(b), can be written as
\begin{equation}\label{eq:H}
\hat{\mathcal{H}}_m = \frac{\mu}{2} \sum_{n=1}^L (-1)^n \hat{\sigma}^z_n - J \sum_{n=1}^{L-1} (\hat{\sigma}_n^{+} \hat{\tau}^+_{n,n+1} \hat{\sigma}^-_{n+1} + \mathrm{h.c.}). 
\end{equation}
We have assumed a system of $L$ matter sites with open boundary conditions. The term proportional to the fermion mass $\mu$ is staggered, since the matter and antimatter components of the original fermion spinor in the continuum are mapped onto the even and odd sublattices respectively. A background of classical positive charges exists on the odd sublattice such that an empty site, encoded by the state $\ket{0}$ (the eigenstate of $\hat{\sigma}^z$ with eigenvalue $+1$), denotes a positive charge for odd $n$ and vacuum for even $n$. Conversely, an occupied site, encoded by the state $\ket{1}$, denotes vacuum for odd $n$ and a negative charge for even $n$. Furthermore, we represent the gauge field pointing to the left as the state $\ket{0}$ and the field pointing to the right as the state $\ket{1}$. The $U(1)$ gauge symmetry is preserved in the QLM with spin-$1/2$ links: any state in the physical Hilbert space, $\ket{\psi_{\mathrm{phys}}}$, follows  Gauss's law $\hat{G}_n\ket{\psi_{\mathrm{phys}}}=0$ defined by the modified symmetry generator
\begin{equation}\label{eq:Gsymm}
 \hat{G}_n = \frac{1}{2}[\hat{\sigma}^z_n - (-1)^n] - \frac{1}{2}(\hat{\tau}^z_{n,n+1} - \hat{\tau}^z_{n-1,n}).
\end{equation}
The first term represents the site charge with the staggered classical background contribution, and the second term is the electric field divergence calculated as the difference between the electric fields on the links connected to site $n$.

%\section{\label{sec:device} }

\begin{figure}
\centering
\includegraphics[width=\linewidth]{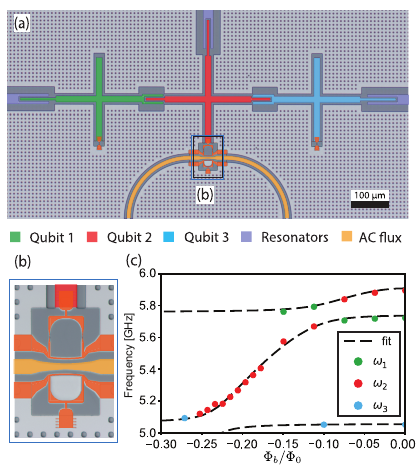}
\caption[ ]{The three-qubit device. (a) Micrograph of the device. Three cross-shaped transmon qubits, with junctions and shunting capacitors, are capacitively coupled to each other. Each qubit is capacitively coupled to a readout resonator. (b) Zoomed-in view of the SQUID of the middle qubit, inductively coupled to the AC flux line. (c) The measured spectrum of single-photon transitions from the ground state, along with the theory fit, as functions of the DC flux bias, $\Phi_b$.}
\label{fig:3qubit}
\end{figure}

We now describe the experimental realization of dynamical gauge fields on our AQS platform (see Appendix~\ref{app:setup} for additional experimental details). Our device is composed of three transmon qubits that have a pairwise capacitive coupling (see Fig.~\ref{fig:LGTs}(c)). Each transmon consists of a Josephson junction and a shunt capacitor. The central transmon is made tunable by replacing the single junction with an asymmetric SQUID. Fig.~\ref{fig:3qubit}(a) shows the central portion of the device which includes the qubits. In addition, the device has three individual resonators, connected to a shared feedline, to allow for multiplexed readout. The SQUID loop is inductively coupled to an on-chip flux line (see Fig.~\ref{fig:3qubit}(b)) which allows us to modulate the SQUID energy to control and activate the interactions parametrically. The transmons are designed at different frequencies such that the transverse interactions are negligible when not activated by parametric pumping. %Each qubit is dispersively coupled to its readout resonator, and the resonators are coupled to a two-port transmission line for multiplexed control and readout. %In the absence of the external flux, 

Treating each transmon in the two-level approximation, the effective Hamiltonian of the system without pumping is
\begin{equation}\label{eq:threequbit}
\begin{aligned}
     \hat{\mathcal{H}}_0 =& \sum_{n=1}^{3} \hbar\omega_n \frac{\hsigma^z_n+1}{2}\\   
     &+ \sum_{n =1}^{2}\hbar\chi_{n,n+1} \,\frac{\hsigma^z_n+1}{2}\frac{\hsigma^z_{n+1}+1}{2},
     \end{aligned}
 \end{equation}
where $\omega_n$ denotes the qubit frequency and $\chi_{n,m}$ the $ZZ$ coupling strength which arises from the cross-Kerr coupling of the transmons. Here, we omit the transverse coupling terms, as discussed above.  

We use an asymmetric SQUID to introduce cubic nonlinearities in the three-qubit system~\cite{sandbo3photon}. An alternative approach based on a distinct device, dubbed the asymmetrically threaded SQUID, is described in Ref.~\cite{lescanne20}. To see the necessity of the asymmetry, we begin from the SQUID Hamiltonian:
\begin{equation}
     \hat{\mathcal{H}}_{\mathrm{sq}} = -E_\mathrm{J} \cos\frac{\pi\hat{\Phi}_{\mathrm{ext}}}{\Phi_0}\cos \hat{\phi}_{2}+\delta E_\mathrm{J} \sin \frac{\pi\hat{\Phi}_{\mathrm{ext}}}{\Phi_0}\sin \hat{\phi}_{2},
     \label{eq:squid8}
\end{equation}
where $E_\mathrm{J}= E_{\mathrm{J},L} + E_{\mathrm{J},R}$, $\delta E_\mathrm{J}= E_{\mathrm{J},L} - E_{\mathrm{J},R}$, $E_{\mathrm{J},i}$ is the Josephson energy of junction $i$ in the SQUID ($i=L,R$), $\hat{\phi}_{2}$ is the SQUID phase operator, $\Phi_0$ is the flux quantum, and $\hat{\Phi}_{\mathrm{ext}} =\hat{\Phi}_{p}(t) + \Phi_{b} $ is the external flux composed of the DC flux bias $\Phi_{b}$ and the AC parametric drive $\hat{\Phi}_{p}(t)$. Here, we work at zero DC bias $\Phi_b = 0$. In the limit of a large-amplitude coherent external drive, we further apply the parametric approximation to represent $\hat{\Phi}_p(t)$ as a classical signal, $\hat{\Phi}_p(t) \rightarrow \alpha_p(t) = A_p \cos(\omega_p t+\varphi)$ where $A_p$, $\omega_p$, and $\varphi$ are the amplitude, frequency, and phase of the parametric drive. Since the qubit modes are mixed owing to strong capacitive coupling at the scale of $100$~MHz, the SQUID is coupled to all modes. We can quantify the contribution of each qubit to the SQUID phase by expressing $\hat{\phi}_{2} = \sum_{n=1}^3 (\lambda_n \hsigma^+_n + \lambda_n^* \hsigma^-_n )$, where the complex coefficients $\lambda_n$ depend on circuit parameters. We expand $\hat{\mathcal{H}}_{\mathrm{sq}}$ in powers of $\hat{\phi}_{2}$ finding
\begin{equation}
     \hat{\mathcal{H}}_{\mathrm{sq}} = -\sum_{k} g_k(A_p) [\sum_{n=1}^3 (\lambda_n \hsigma^+_n + \lambda_n^* \hsigma^-_n )]^k,
     \label{eq:hsq_pow}
\end{equation}
where the coefficients $\{g_k\}$ are functions of $A_p$. In the interaction picture of $\hat{\mathcal{H}}_0$, different terms in Eq.~\eqref{eq:hsq_pow} acquire different time dependencies and can thus be selectively activated by parametrically driving the SQUID at the corresponding frequencies. In particular, the resonance frequency $\omega_{001 \rightarrow 110} = (E_{110}-E_{001})/\hbar$ corresponds to the transition between the states $\ket{001}$ and $\ket{110}$ with energies $E_{001}=\hbar\omega_3$ and $E_{110}=\hbar(\omega_1 + \omega_2 + \chi_{1,2})$ respectively. Driving at $\omega_p \approx \omega_{001 \rightarrow 110}$ yields a time-independent three-qubit interaction. Applying the rotating-wave approximation, we find $ -J(A_p) \hsigma^+_1 \hsigma^+_2 \hsigma^-_3 +\mathrm{h.c.}$, where the effective 3-qubit interaction strength, $J(A_p)$, is a function of $A_p$. Arising from the cubic term in Eq.~\eqref{eq:hsq_pow}, this interaction can only come from the $\delta E_\mathrm{J} \sin \hat{\phi}_{2}$ term, which is due to the SQUID asymmetry. To the leading order in $A_p$, $J(A_p) \propto \delta E_\mathrm{J} A_p$ (see Appendix~\ref{app:theory}).

Finally, we rewrite our qubit Hamiltonian using the LGT notation, so that the system now represents a lattice with two sites and one link (Fig.~\ref{fig:LGTs}(b)). The qubits 1 and 3 represent matter sites 1 and 2, while the gauge field of the $1,2$ link is represented by the state of qubit 2.
%an empty site is encoded as the qubit ground state $\ket{0}$ and an occupied site as the excited state $\ket{1}$. In the case of spin-$1/2$ gauge fields, we represent the field pointing to the left as the qubit ground state $\ket{0}$ and the field pointing to the right as the excited state $\ket{1}$.
The states $\ket{001}$ and $\ket{110}$ (Fig.~\ref{fig:LGTs}(b)) form the set of possible system states that satisfy Gauss's law, spanning the $U(1)$ gauge-invariant subspace defined by the symmetry generator in Eq.~\eqref{eq:Gsymm}. Choosing the rotating frame with frequencies $\epsilon_{001}=E_{001}-\hbar\delta\omega/2$ and $\epsilon_{110}=E_{110}+\hbar\delta\omega/2$, where $\delta\omega=\omega_p-\omega_{001 \rightarrow 110}$ is the detuning between the parametric drive frequency and the resonance frequency, we arrive at the system Hamiltonian in the interaction picture in this gauge-invariant subspace:
\begin{equation}\label{eq:h_int_3q}
\begin{aligned}
     \hat{\mathcal{H}}_{\mathrm{int}}  \approx & \frac{\mu(\omega_p)}{2}\sum_{n= 1,2} (-1)^n \hat{\sigma}^z_n\\
     & - J(A_p) ( \hsigma^+_1 \hat{\tau}^+_{1,2} \hsigma^-_2 +\mathrm{h.c.}),
\end{aligned}
\end{equation}
where $\mu(\omega_p) = -\hbar\delta\omega/2=-\hbar(\omega_p-\omega_{001 \rightarrow 110})/2 $. It is worth noting that we can control the effective site energy through the detuning of the parametric drive even in the presence of a nearest-neighbor $ZZ$ coupling. In Appendix~\ref{app:map}, we further demonstrate this tunability in a longer chain of 4 matter sites and 3 gauge sites.

\section{\label{sec:characterize} Device characterization}

We characterize the system by performing spectroscopy on the qubit frequencies, $\omega_n$,  as functions of the flux bias $\Phi_b$. In addition, we perform two-tone spectroscopy to identify the higher energy levels (two-photon states) at the desired operating point $\Phi_b =0$, which allows us to estimate the additional Hamiltonian parameters $\chi_{i,j}$. The measured spectrum, along with the fit using our circuit model, is shown in Fig.~\ref{fig:3qubit}(c). The extracted circuit parameters are provided in Table~\ref{tab:fittingParams2}, and further details on device characterization are given in Appendix~\ref{app:characterize}.

% \setlength{\tabcolsep}{6pt}
% \begin{table}
%     \centering
%     \begin{tabular}{|c | c c c|}
%     \hline
%          Qubit & 1 & 2 & 3  \\
%          \hline 
%          $\omega^{q}_{01} $ [GHz] & 5.725& 5.910&  5.055\\

%          Qubit capacitance, $C_n$ [fF] & 103& 111& 101\\
%          \hline \hline
%         Qubits $n$-$m$  &  1-2 & 1-3 & 2-3  \\
%         \hline
%        Coupling capacitance, $C_{nm}$ [fF]&  2.33& 0.58& 4.25\\
%         \hline 
%     \end{tabular}
%     \caption{Measured qubit frequencies and parameters extracted by fitting the three-qubit model to the spectrum as a function of the DC flux bias.}
%     \label{tab:fittingParams2}
% \end{table}

\setlength{\tabcolsep}{6pt}
\begin{table}
    \centering
    \begin{tabular}{|c | c c c|}
    \hline
         Qubit & 1 & 2 & 3  \\
         \hline 
         $\omega_{n}/2\pi $ [GHz] & 5.7279& 5.9098&  5.0538\\

         $(E_{\mathrm{C}}/\hbar)/2\pi$ [MHz]& 183& 165& 184\\
         \hline \hline
        Qubits $n$-$m$  &  1-2 & 1-3 & 2-3  \\
        \hline
         $(g_{nm}/\hbar)/2\pi$ [MHz]&  63& 18& 108\\
        \hline 
    \end{tabular}
    \caption{Qubit parameters. Measured qubit frequencies $\omega_n$ at zero DC flux bias, charging energies $E_{\mathrm{C}}$, and capacitive coupling constants $g_{nm}$ (defined in Eq.~\eqref{eq:g_jk_capacitive}). To extract $E_{\mathrm{C}}$ and $g_{nm}$, we fit the three-qubit circuit model to all of the following measured data: the $\omega_n$ as functions of the DC flux bias $\Phi_b$ (Fig.~\ref{fig:3qubit}(c)); two-photon energy levels at $\Phi_b=0$ (Table~\ref{tab:TransitionFreq}); and the three-qubit interaction strength and resonance frequency at $\Phi_b=0$ as functions of the parametric drive strength, $A_p$ (see Fig.~\ref{fig:ELevels}(c)).}
    \label{tab:fittingParams2}
\end{table}

\begin{figure*}
    \centering
    \includegraphics[width=\linewidth]{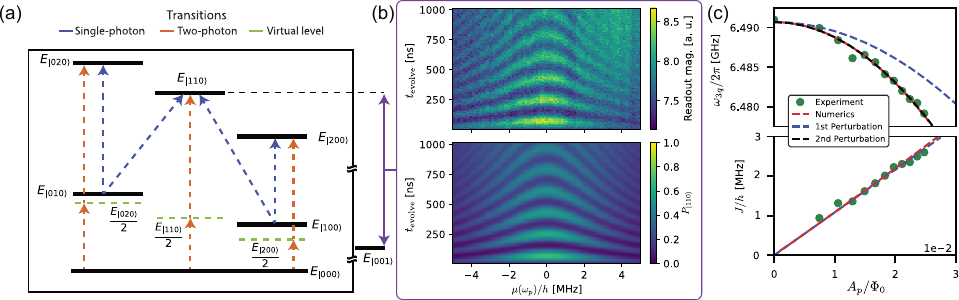}
    \caption{Characterizing the three-body interaction. (a) Validation of the $\ket{110}$ state. We prepare the system in $\ket{010}$ or $\ket{100}$ as two possible single-photon paths to $\ket{110}$. We use low-amplitude control pulses to suppress multiphoton transitions. Although the two-photon states  $\ket{020}$ or $\ket{200}$ can be accessed with single-photon transitions (blue) via one of the paths, the only state accessible via single-photon transition (blue) in both paths is the state $\ket{110}$. After state identification, we use a stronger drive at half the transition frequency to drive the higher states via two-photon processes (orange). (b) 2D Rabi-style chevron of three-body interaction between $\ket{110}$ and $\ket{001}$ as we measure the $\ket{110}$ state by the readout magnitude at resonator 2 (top) and population extracted from a fit to the cavity-Bloch equations (bottom). (c) Measured interaction strength, $J$, and resonance frequency, $\omega_{3q}$, (green markers) as functions of the parametric pulse amplitude, $A_p$, together with first-order (blue) and second-order (black) perturbation theory and brute-force numerical calculations (red) using model parameters from the fit in Fig.~\ref{fig:3qubit}(c) and Table~\ref{tab:fittingParams2}. We see that the second-order perturbation theory agrees well with the brute-force calculations. We note that first-order perturbation theory agrees well for $J$ while producing significant deviations for $\omega_{3q}$. }  
    \label{fig:ELevels}
\end{figure*}

\begin{table}
    \centering
    \begin{tabular}{|c | c c c c|}
    \hline
         \multirow{2}{*}{\diagbox[width=5em,height=2\line]{To}{From}}& \multirow{2}{*}{$\ket{000}$} & $\ket{100}$ & $\ket{010}$ & $\ket{001}$  \\
               &             & {\color{blue}$5.7279$} & {\color{blue}$5.9098$}  & {\color{blue}$5.0538$}    \\
         \hline 
         \multirow{2}{*}{$\ket{200}$} & {\color{orange}$5.6352$} & {\color{blue}$5.5435$} & \multirow{2}{*}{$-$} &  \multirow{2}{*}{$-$} \\ 
                     & $(11.2704)$ & $(11.2714)$ &    &    \\
         \hline 
         \multirow{2}{*}{$\ket{110}$} & {\color{orange}$5.7742$} & {\color{blue}$5.8094$} & {\color{blue}$5.6345$} & {\color{violet}$6.4907$} \\ 
                     & $(11.5484)$ & $(11.5373)$ & $(11.5443)$   & $(11.5445)$   \\
         \hline 
         \multirow{2}{*}{$\ket{020}$} & {\color{orange}$5.8582$} & \multirow{2}{*}{$-$} &  {\color{blue}$5.806$} & \multirow{2}{*}{$-$} \\ 
                     & $(11.7164)$ &  &  $(11.7158)$  &    \\
         \hline 
    \end{tabular}
    \caption{Measured transition frequencies in GHz from the ground state and the single-photon states to the three two-photon states $\ket{200}$, $\ket{110}$ and $\ket{020}$ involved in the validation of $\ket{110}$. These states are measured through single-photon transitions (blue) and two-photon transitions (orange). For each transition, we also show the energy of the final state calculated by summing measured transition frequencies in parentheses below the transition frequency. The only three-photon transition frequency (violet) we have measured is $\omega_{001 \to 110}$. Transition frequencies from the ground state to the single-photon states are also given in the first row.}% $\ket{100}$, $\ket{010}$ and $\ket{001}$
    \label{tab:TransitionFreq}
\end{table}

Choosing the proper parametric frequency requires us to identify the target transition $\omega_{001 \rightarrow 110}$, which, in our encoding, represents the hopping of the charge accompanied by the change in direction of the field. In this system of strongly coupled qubits, we adopt the following procedure to map out the energy levels of interest (see Fig.~\ref{fig:ELevels}(a)). First, we calibrate the amplitude of the spectroscopy pulse to ensure it is weak enough to not excite multiphoton transitions. Then, we climb the energy ladder via two consecutive single-photon transitions as follows. We prepare the system in the single-photon state $\ket{010}$  or $\ket{100}$ by applying the appropriate $\pi$-pulse. We then perform a Rabi-style experiment in the frequency range where we expect the transition to the two-photon state $\ket{200}, \ket{020}$ or $  \ket{110}$.  When the pulse frequency is in resonance with a single-photon transition, we observe Rabi-like oscillations between the single-photon state and the two-photon state. 

The experiment is done twice for each two-photon state, climbing the ladder through the two possible paths, which transit through either $\ket{100}$ or $\ket{010}$. Each path connects to two states via a single-photon transition, e.g.,  $\ket{100}$ connects to $\ket{200}$ and $\ket{110}$. % and  $\ket{010} \rightarrow \ket{020}, \ket{110}$. 
However, the transitions $\ket{100} \rightarrow \ket{020}$ and  $\ket{010} \rightarrow \ket{200}$ are higher-order processes involving the annihilation of a photon in one qubit and the creation of two photons in another. Accordingly, they are suppressed under the weak driving conditions used for this procedure. Hence,  this allows us to determine the target state $\ket{110}$,  since it is the only state accessible via single-photon transitions through both paths. That is, we identify $\ket{110}$ by finding the transition frequencies which satisfy $\omega_{000\to 010} + \omega_{010\to 110} = \omega_{000\to 100} + \omega_{100\to 110}$. As a further confirmation, we probe the two-photon states identified through the serial single-photon transitions by directly probing the two-photon transitions at, e.g., $\omega_{000\to 110}/2$. %half the transition frequencies starting from the ground state, i.e.,  $\ket{000} \rightarrow \ket{200}, \ket{020}, \ket{110}$.

% \begin{figure*}
%     \centering
%     \includegraphics[width=\linewidth]{Figures/Fig3_v2.pdf}
%         \caption{(a) Validation of the $\ket{110}$ state. Solid black lines are measured energy levels. We prepare the system in $\ket{010}$ or $\ket{100}$ as two possible paths to $\ket{110}$. We use low-amplitude control pulses to suppress multiphoton transitions. Although the two-photon states  $\ket{020}$ or $\ket{200}$ can be accessed via single-photon transitions via one of the paths, the only state accessible via single-photon transition in both paths is the state $\ket{110}$. The dashed lines depict the two-photon frequencies in the absence of cross-Kerr couplings. After state identification, we use a stronger drive at half the transition frequency to drive the two-photon states via a two-photon process (orange). (b) 2D Rabi-style chevron of three-body interaction between $\ket{110}$ and $\ket{001}$ as we measure the $\ket{110}$ state by readout at resonator 2. (c) Extracted interaction strength $g_{\mathrm{eff}}$ and resonance frequency $\omega_{3q}$ (green markers) as functions of the parametric pulse amplitude, together with predictions of perturbation theory (black line) and results extracted from brute-force numerical calculations (red line) using model parameters from the spectrum fit in (a) and Fig.~\ref{fig:3qubit}(c).}  
%     \label{fig:ELevels}
% \end{figure*}

\section{\label{sec:results} Realization of three-body interaction}

\begin{figure}
    \centering
    \includegraphics{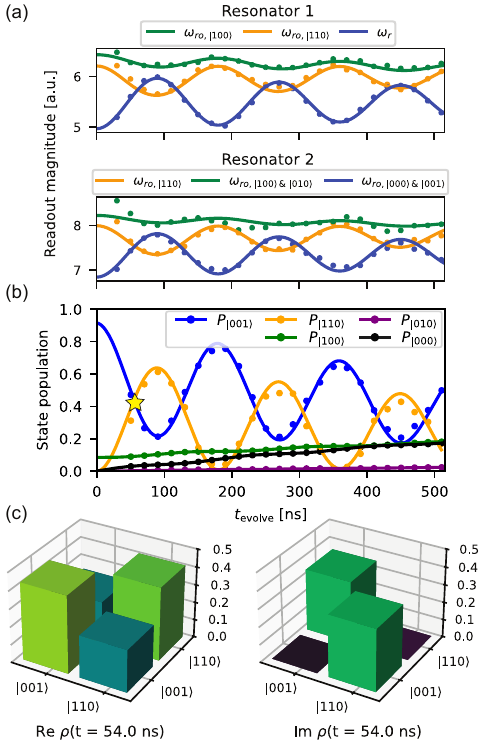}
    \caption{Coherent oscillations driven by three-body interaction. (a) Magnitudes of the measured response for resonators 1 and 2, each probed at multiple different readout frequencies (markers) as the system evolves under the three-body Hamiltonian for the evolution time, $t_{\mathrm{evolve}}$.  We probe each resonator at multiple frequencies so that we can extract the state of the qubits despite the significant cross-coupling between the resonators and the strong coupling of the  qubits.  The readout frequencies include the (ground-state) resonator frequency, $\omega_r$, and multiple state-shifted resonator frequencies, $\omega_{ro, \ket{s}}$, corresponding to different qubit states $\ket{s}$. The theory fits (solid lines) are obtained using the cavity-Bloch equations, which describe the dynamics of the coupled qubit-resonator system and take into account five qubit states: $\ket{001}, \ket{110}, \ket{100}, \ket{010}$ and $\ket{000}$. (b) Normalized state populations as functions of $t_{\mathrm{evolve}}$ (markers), obtained by rescaling the data in panel (a) using parameters from the fit to cavity-Bloch equations (solid lines; see main text for details). There are coherent oscillations between the initial state $\ket{001}$ (blue) and the target state $\ket{110}$ (orange), and also relaxation to lower energy states $\ket{100}$ (green), $\ket{010}$ (purple) and $\ket{000}$ (black). (c) Real and imaginary parts of elements of the reconstructed density matrix, extracted at $t_{\mathrm{evolve}}=54$~ns, marked with a star symbol in panel (b). The strong off-diagonal matrix elements, e.g.  ${\ket{110}\bra{001}}$, highlight the coherence of the realized superposition state.}
    \label{fig:Measurements}
\end{figure}

\begin{figure}
\centering
\includegraphics{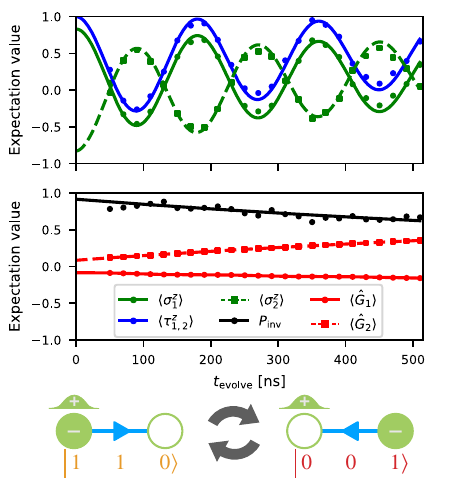}
\caption{Evolution of site spins and gauge-invariant quantities. (Upper panel) Expectation value of the spin-$z$ operator for each matter (green circles/squares) and gauge (blue) qubit. (Lower panel) The total population of the gauge-invariant sector $P_{\text{inv}}=P_{\ket{110}}+P_{\ket{001}}$ (black) and the expectation values of the symmetry generators $\langle\hat{G}_{i}\rangle$ (red circles/squares). Markers (lines) are calculated using the measured (fit) populations in Fig.~\ref{fig:Measurements}. $\langle \hat{\sigma}^z_1 \rangle$, $\langle \hat{\tau}^z_{1,2} \rangle$ and $\langle \hat{\sigma}^z_2 \rangle$ all exhibit coherent oscillations consistent with the dynamics in the gauge-invariant sector. The strongly suppressed oscillations of $P_{\text{inv}}$ and $\langle\hat{G}_{i}\rangle$ indicate that the system satisfies Gauss's law within the computational subspace and the qubit lifetime. }%It decays exponentially due to relaxation out of the computational states, though we see it matches the peaks of $\ket{110}$ and $\ket{001}$ populations.
    \label{fig:LGTsSimulation}
\end{figure}

To demonstrate the three-body interaction, we initialize the system in the state $\ket{001}$  by applying a $\pi$ pulse on qubit 3. Then, we activate the three-body interaction for a time $t_{\mathrm{evolve}}$ (the ``evolution time'') by applying a parametric pulse at approximately $\omega_{001 \rightarrow 110}$. The three-body interaction simultaneously annihilates (creates) a single excitation in qubit 3 and creates (annihilates) a pair of excitations in qubits 1 and 2, driving coherent oscillations between the states $\ket{110}$ and $\ket{001}$. In Fig.~\ref{fig:ELevels}(b), we show the 2D Rabi-like chevrons produced by repeating the experiment while varying $t_{\mathrm{evolve}}$ and $\omega_p$. For this data, we used resonator 2 for readout.  We measure the resonance frequency, $\omega_{3q}$, and the interaction strength, $J$, as functions of the pulse amplitude, $A_p$, achieving an interaction strength up to $J/2\pi\hbar = 3$~MHz (see Fig.~\ref{fig:ELevels}(c)). 
%and fit perturbation theory to first order in $\alpha_p$, $\alpha_p \propto \langle 110|\sin \hat{\phi}_{2}| 001 \rangle$  . 

For each value of $A_p$, we measure $\omega_{3q}$ by finding the center of the Rabi chevrons. We see that $\omega_{3q}$ shifts downward from $\omega_{001 \rightarrow 110}$ as $A_p$ increases.  This shift is only partly explained by the first-order perturbation theory commonly employed in literature~\cite{reagor18, noh23, yang23}, where the curvature of the energy bands as a function of flux gives the effective shift of the qubit energies to first order in the drive Hamiltonian (and thus to the second order in $A_p$). Because our system is strongly coupled and $A_p$ is relatively large, we needed to include an additional Bloch–Siegert-type shift that arises in second-order perturbation theory (see Appendix~\ref{app:numerics}). With that, we were able to obtain quantitative agreement, as shown in Fig.~\ref{fig:ELevels}(c), between the analytical results, experimental measurements, and brute-force numerical calculations.

To perform a dispersive readout of the state of the qubits, we apply a readout pulse near resonance with each resonator and measure the quadrature amplitudes transmitted through the feedline as functions of time and readout frequency. These measurements are fit to the cavity-Bloch equations~\cite{Bianchetti09}, a set of differential equations for the time-dependent expectation values of the qubit operators and the resonator field (see Appendix~\ref{app:extract}). Given that the qubit lifetimes are comparable to the resonator decay time, the cavity-Bloch equations are essential for capturing the interplay between the resonator and the qubit state, in particular, accounting for the qubit population decay during the readout.

First, to characterize the resonators, we keep the qubits in the $\ket{000}$ state and fit the dynamics of the resonators under the readout pulses to obtain the internal and external loss rates. We then prepare the system in each of the four states $\ket{001}$, $\ket{110}$, $\ket{100}$ and $\ket{010}$, and determine the corresponding dispersive shifts of the resonator frequencies, again by fitting the readout signal to the cavity-Bloch equations. Owing to the strong coupling between qubits 1 and 2, resonators 1 and 2 exhibit strong dispersive shifts for both $\ket{100}$ and $\ket{010}$. 

Having characterized the resonators and the dispersive shifts of all relevant qubit states, we are now ready to fit the evolution of the full qubit-resonator system under the three-body Hamiltonian. We assume the qubit state before the $\pi$ pulse on qubit 3 is an incoherent mixture of $\ket{000}$, $\ket{001}$, $\ket{100}$ and $\ket{010}$, where all single-photon populations are thermal. We further assume that the $\pi$ pulse on qubit 3 fully converts the $\ket{000}$ population to $\ket{001}$ before the three-body interaction is activated. The response of the resonators for different $t_{\mathrm{evolve}}$ and readout pulse frequencies is then fit to a model with five qubit levels: $\ket{000}$, $\ket{001}$, $\ket{100}$, $\ket{010}$ and $\ket{110}$ (see Appendix~\ref{app:extract}).

In Fig.~\ref{fig:Measurements}(a), we show the measured average resonator quadrature amplitudes as functions of $t_{\mathrm{evolve}}$, alongside their fits. The measurements are repeated at different readout frequencies, since at any given frequency the resonators have different sensitivities to different states of the qubits. In particular, we measure at the $\ket{110}$ readout frequencies on both resonators to extract the coherent evolution. We also measure resonator 1 at the $\ket{100}$ readout frequency, and resonator 2 at $\ket{100}$ and $ \ket{010}$ readout frequencies; these measurements tell us about relaxation (decoherence) during the evolution. Finally, we measure all resonators at their center frequencies for information of the states $\ket{000}$ and $\ket{001}$. 

We show the extracted system populations in Fig.~\ref{fig:Measurements}(b). The populations $P_{\ket{100}}, P_{\ket{010}}$ and $P_{\ket{000}}$ are directly taken from the theory fit, while the populations $P_{\ket{001}}$ and $P_{\ket{110}}$ are obtained by rescaling the corresponding readout signals in Fig.~\ref{fig:Measurements}(a) using scale factors extracted from the fit~\footnote{Note that rescaling the readout signals at the ground-state resonator frequency yields $P_{\ket{001}}+P_{\ket{000}}$, and we need to further subtract the fit values of $P_{\ket{000}}$ to obtain $P_{\ket{001}}$}. During the coherent oscillation in the subspace spanned by $\ket{110}$ and $\ket{001}$, the extracted expectation value of the off-diagonal element of the density matrix, $\ket{110}\bra{001}$, clearly indicates a coherent superposition when half of the population is transferred to $\ket{110}$ at $t_{\mathrm{evolve}}=54$~ns (Fig.~\ref{fig:Measurements}(c)).

The oscillations decay as a result of dephasing and energy relaxation. We observe an increase in the populations $P_{\ket{100}}, P_{\ket{010}}$ and $P_{\ket{000}}$ outside the coherent subspace due to standard single-qubit energy relaxation. We can extract three decay rates: $1/\gamma_{110\to010}=6.6\pm1.5\,\mu$s, $1/\gamma_{110\to100}=1.561\pm0.092\,\mu$s, and $1/\gamma_{001\to000}=1.28\pm0.33\,\mu$s. These agree qualitatively with the $T_1$ lifetimes of the three qubits in Table~\ref{tab:qubit_characterize} in Appendix~\ref{app:characterize}, which have been separately extracted from single-qubit measurements. Discrepancies may relate to different transition matrix elements as well as frequency shifts of the two-photon states compared to the single-photon states. % That is, the relaxation process is happening at a somewhat different frequency in the two cases. % In this interaction, a single excitation in qubit 3 is annihilated, and a pair of excitations are created in qubit 1 and 2 simultaneously. 

\begin{figure*}
\centering \includegraphics[width = \linewidth]{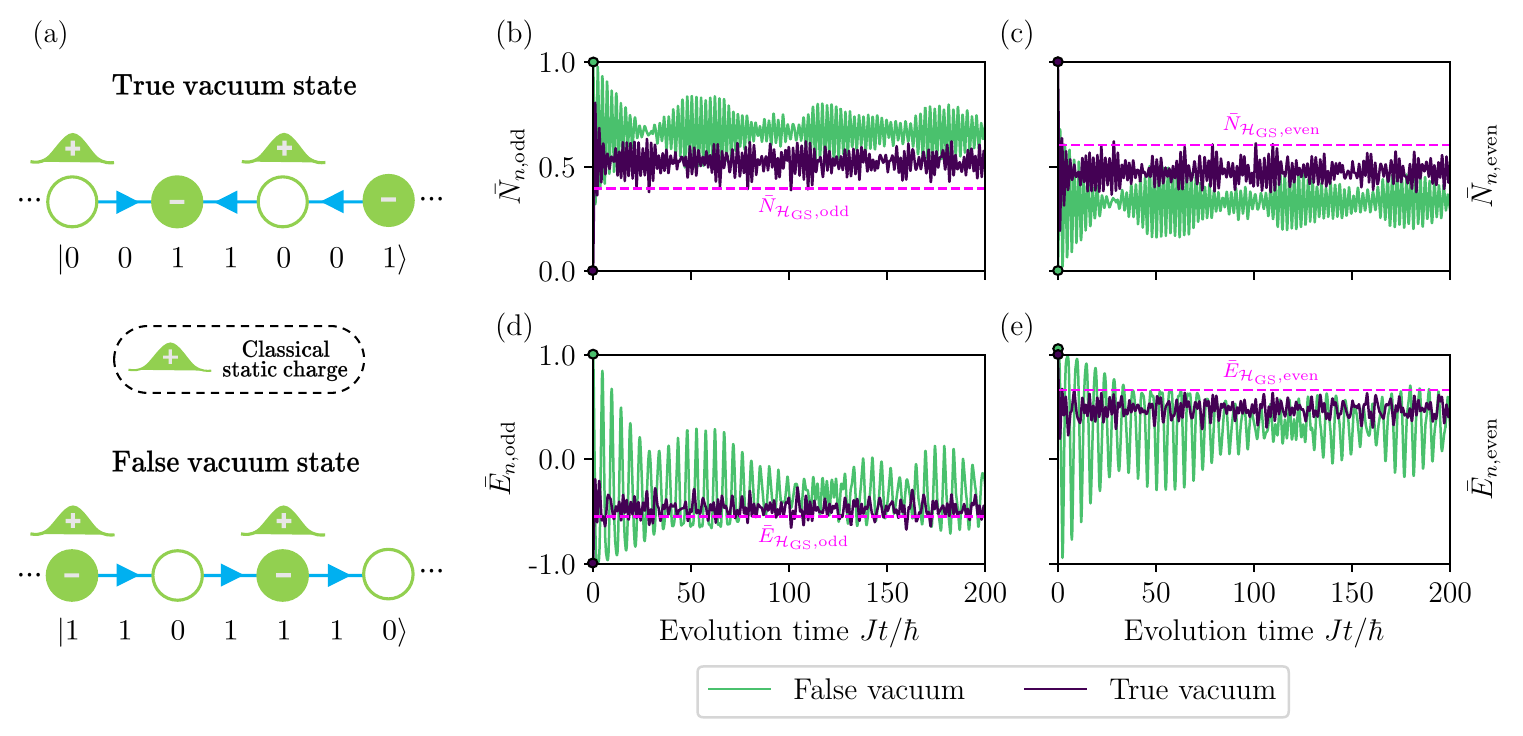}
\caption{Recovering the symmetry-preserving true vacuum configuration with a gauge-invariant time evolution from a false vacuum state. Our model assumes a classical background, representing a structured vacuum state, akin to the Dirac sea in quantum field theory. This background functions as a stable reference state, while our qubit dynamics track deviations and responses to perturbations from this background. (a) Representation of the two types of central states for the false vacuum decay of the system in the strong coupling limit $|\mu / J |\gg 1$. The total net electric flux $\hat{E} = \sum_{n = 1}^{L-1} \hat{E}_n$ acts as the order parameter of two phases, distinguished by whether the parity and charge conjugation symmetries are broken or not. The upper state is the true vacuum of the gauge-invariant system when $\mu / J \gg 1$. The bottom state is one of the two false vacua of the system at $\mu / J \gg 1$. The false vacua form a two-fold degenerate phase where the parity and charge conjugation symmetries are broken, having a nonzero net electric flux in two possible directions. Here we represent the state with the flux pointing to the right. The other configuration consists of the flux pointing to the left. For clarity, the states have been represented on a 4-site 1D lattice, even though the simulations are carried out on a 12-site lattice. (b)/(c) Comparison of the particle number $\hat{N}_n$ on the odd (left) and even (right) matter sites, beginning from either a false vacuum or the true vacuum. (d)/(e) Comparison of the electric field $\hat{E}_n$ per odd (left) and even (right) link, beginning from either a false vacuum or the true vacuum. We observe that the evolution from the false vacuum state shows much bigger oscillations in time, i.e., it is more unstable than the true vacuum evolution. After some initial time evolution, the averaged values in the false vacuum converge towards the true vacuum ones. The purple and green points indicate the starting values of the magnitudes in the false and true vacuum schemes respectively. Everything is also compared to the calculated magnitudes for the ground state of the Hamiltonian containing just the three-body interaction (called $\mathcal{H}_{\rm GS}$ here), represented with a dashed fuchsia line. The averaged magnitudes have been calculated with the values of the three central odd and even sites/links acting as the bulk, leaving the rest as the environment. The simulation is done with a 12-site lattice for a total evolution time $Jt/\hbar = 200$, and a Hamiltonian with site energy $\mu/J = 0$.}
    \label{fig:NandE_oddeven_other}
\end{figure*}
%

% To fit the evolution of the full qubit-resonator system under the three-body Hamiltonian, we assume the qubit state before the $\pi$ pulse on qubit 3 is an incoherent mixture of $\ket{000}$, $\ket{001}$, $\ket{100}$ and $\ket{010}$, where all single-photon populations are thermal. We further assume that the $\pi$ pulse fully converts the $\ket{000}$ population to $\ket{001}$. In Figure.~\ref{fig:Measurements}(a), we show the theory and experiment of the resonator averaged quadrature amplitudes as a function of evolution time fit across different readout frequencies and various coherent evolution times. In the extracted system population shown in Fig.~\ref{fig:Measurements}(b), we see the coherent oscillation between the population of $\ket{110}$ and $\ket{001}$, as well as decay to lower states $\ket{100}$ and $\ket{000}$. In addition, the extracted expectation value of the off-diagonal element of the density matrix, $P_{\ket{110}\bra{001}}$, clearly indicates a coherent superposition when half of the population is transferred to $\ket{110}$ at evolution time of 49 ns. 

In Fig.~\ref{fig:LGTsSimulation}, we further illustrate how the demonstrated three-qubit interaction realizes the matter-gauge interaction of a  2-site $U(1)$ LGT while maintaining Gauss's law. Initially, we started with state $\ket{001}$, encoding an excitation at matter site 2 with the electric field at the link pointing to the left, satisfying Gauss's law. Once the three-body interaction is activated, the system starts evolving toward state $\ket{110}$, simulating the matter excitation hopping from site 2 to site 1 while simultaneously flipping the direction of the electric field to the right, such that the system preserves Gauss's law in Eq.~\eqref{eq:Gsymm}. 

To quantify this, we calculate the population in the gauge-invariant subspace $P_{\text{inv}}=P_{\ket{110}}+P_{\ket{001}}$, as well as the expectation values of the symmetry generators, $\langle\hat{G}_1\rangle =  \frac{1}{2}\langle\hat{\sigma}^z_1\rangle - \frac{1}{2}\langle\hat{\tau}^z_{1,2}\rangle$ and $\langle\hat{G}_2\rangle =  \frac{1}{2}\langle\hat{\sigma}^z_2\rangle + \frac{1}{2}\langle\hat{\tau}^z_{1,2}\rangle$. (We have removed the constants $\pm 1/2$ in Eq.~\eqref{eq:Gsymm} to account for the fact that both matter sites are on the open boundaries of this system.) The high-contrast oscillations in computational states of the LGT are strongly reduced in the gauge-invariant sum $P_{\text{inv}}$ and the expectation values of the symmetry generators $\langle\hat{G}_{i}\rangle$, indicating that Gauss's law is preserved by the system's dynamics. The remaining exponential decay is due to the relaxation of the system out of the computational subspace. This illustrates that we achieved gauge-invariant dynamics using the three-body interaction in our three-qubit system.

\section{\label{sec:time_evolution} Numerical simulations of larger systems}

Having demonstrated the basic building block of a lattice gauge simulator, we now want to explore the potential of the realized Hamiltonian in larger simulations. Quantum simulators of gauge fields offer a unique opportunity to study the real-time evolution of gauge-invariant models, an important class of quantum systems. They allow for the exploration of numerous inaccessible scenarios through conventional techniques. One example is the process of false-vacuum decay, which can be closely tracked and replicated step-by-step on these quantum processors. 

Broadly, this scenario examines a gauge-invariant model initially residing in a metastable state called the ``false vacuum". In our model, this state corresponds to a local minimum of the system in the strong coupling limit $|\mu / J |\gg 1$. While the false vacuum is a metastable configuration in this limit, it is not globally stable. When the gauge-invariant interaction is switched on, the true vacuum configuration appears in the evolved quantum state. A transition between these vacuum states can occur through a process wherein the field can ``tunnel" from the false vacuum to the true vacuum. This tunneling mechanism is analogous to the well-known quantum-mechanical tunneling effect but for quantum fields. 

In this section, we use a classical numerical simulation to study the phenomenon of false vacuum decay in a 12-site 1D lattice described by our realized Hamiltonian, Eq.~\eqref{eq:H}. Our model tracks the field's evolution throughout this transition, dynamically describing the false vacuum decay process. This process has far-reaching implications in cosmology as it may have played a critical role in shaping the early universe. Understanding the parameters and behaviors that dictate the false vacuum decay allows us to explore scenarios in which the universe transitions from a metastable configuration to a more stable one. Our model’s tracking of field evolution offers a detailed quantitative framework to explore both the theoretical underpinnings and potential observable consequences of vacuum decay, contributing valuable insights into fundamental physics and cosmology.

In the strong-coupling limit $|\mu / J | \gg 1$, not yet influenced by the three-qubit interaction, our gauge-invariant model Hamiltonian, Eq.~\eqref{eq:H}, along with Gauss's law, Eq.~\eqref{eq:Gsymm}, has three configurations that potentially minimize the energy (see Fig.~\ref{fig:NandE_oddeven_other}(a), with more detail in Appendix~\ref{app:string}). One of these states preserves parity and charge conjugation symmetries, and the other two break them. For large positive $\mu$, they are identified with the true vacuum and the false vacua, respectively. 

We simulate a 12-site lattice so that, by beginning with a false vacuum configuration and then undergoing gauge-invariant evolution driven only by the three-qubit interaction $(\text{site energy } \mu / J = 0)$ that couples matter and gauge degrees of freedom, the system progressively transitions toward a configuration resembling the true vacuum. This can be seen in Fig.~\ref{fig:NandE_oddeven_other} where the time evolution of the particle number $\hat{N}_n = ( \mathbb{1} - \hat{\sigma}^z_n )/2 $ per matter site and the electric field $\hat{E}_{n} = -\hat{\tau}^z_{n,n+1} $ per gauge link is computed. Due to the finite size of the string, the values in the central sites and links are averaged as the most representative of the bulk behavior of an infinite system. The specific structure of the Hamiltonian distinguishes the values of these observables on even and odd sites, which evolve toward a configuration with zero net electric field, corresponding to the true vacuum. The time evolution has been done for $\mu / J = 0$ as the effect of the three-body term is more explicit in this regime. Appendix~\ref{app:string} shows similar computations for a range of values of $\mu / J \neq 0$).

%As we have a finite string we choose the central sites/links as the most representative of the infinite system behavior. The specific form of the Hamiltonian clears the different values of these observables on even and odd sites, evolving in time to a configuration with zero net electric field i.e. the true vacuum.

\section{\label{sec:discussion} Discussion}

This paper demonstrates a native three-body interaction as a building block for simulating dynamical gauge fields in LGTs. The building block consists of two matter sites connected by a gauge field link, where all three are encoded as qubits. Our approach leverages parametrically activated nonlinear interactions using a tunable link qubit capacitively coupled to matter qubits at two ends. The interaction strength is controlled by applying a resonant pulse with appropriate amplitude. We observe the time evolution of the dynamical gauge field and matter sites that maintain the gauge symmetry and, as a result, satisfy Gauss's law. The experimental results showcase a promising pathway toward scalable quantum simulators for exploring nonperturbative regimes of LGTs.

Various types of three-qubit interactions and gates have been realized previously in different contexts. Ref.~\cite{menke22} implements a static, longitudinal ($ZZZ$) three-qubit interaction between flux qubits via a common flux-tunable coupler. Ref.~\cite{liu20} perturbatively creates an effective three-qubit chiral spin interaction for pairwise coupled transmon qubits in a loop architecture by periodically modulating the qubit frequencies. In the trapped ion platform, Ref.~\cite{katz23} uses pulse sequences (displacement, squeezing, and $\pi$ pulses) to realize effective transverse $XXX$ and $XXXX$ interactions, which are generally accompanied by lower-order $XX$ terms.  In addition, several works implement three-qubit gates involving single-photon transitions. By using simultaneous two-qubit drives instead of a single three-qubit pulse, Refs.~\cite{kim22,warren23,itoko24} realize different three-qubit gates on fixed-frequency transmons: the $i$Toffoli gate, the controlled-CPHASE-SWAP gate, and the three-qubit parity gate, respectively. Ref.~\cite{roy20} implements controlled-controlled-rotation gates on three strongly cross-Kerr-coupled transmons. Finally, Refs.~\cite{fedorov12,reed12} implements the Toffoli gate by utilizing excited states outside the computational space. 

These interactions are inadequate for U(1) LGTs because they do not generate the required hopping-like interaction or they introduce additional terms that break the gauge invariance. In our work, we implement a transverse three-qubit interaction that allows three-photon transitions, where an excitation is annihilated in one qubit and a pair of excitations is created in two others. The dynamical evolution of all three qubits enables encoding an excitation hopping between qubit 1 and qubit 3, while the gauge dynamics are explicitly represented by simultaneously flipping the state of qubit 2. Importantly, interaction is realized natively at the Hamiltonian level using one single parametric pulse and is particularly well-suited for the AQS of models with dynamical gauge fields. 

Scaling up the number of qubits is essential for simulating more complex quantum systems. In our architecture, qubits are designed to be at distinct frequencies separated by many linewidths ($\gg10$). This naturally allows for frequency multiplexing both in the readout and control lines.  Multiplexing allows us to add more qubits to our system to increase the number of lattice sites with minimal addition of cryogenic complexity. It has been demonstrated that up to 10 qubits can be read out simultaneously without significant degradation in readout fidelity~\cite{song24}. %~\cite{karamlou24, heinsoo18}.
Further, the qubits are unevenly spaced in frequency to allow for selective control of the parametric coupling. In particular, each building block can be designed with a unique resonant frequency for the three-body interaction term, minimizing unwanted crosstalk and interference between qubits~\cite{reagor18,busnaina24}. %[ Chalmers University and MIT]. 
In addition, the move to a flip-chip design will facilitate the transition from 1D to 2D architectures~\cite{rosenberg17,kosen22}, with control and readout lines placed in a separate layer from the qubits. This arrangement paves the way for simulating 2D U(1) LGT, a classically challenging regime where interesting physical phenomena can be explored.

Our approach offers a unique potential to simulate nontrivial physical phenomena, such as string breaking in LGTs, with a relatively modest number of qubits. In a 1D lattice, string breaking can be observed with a few qubits, provided the coupling strength is sufficiently large compared to the coherence times~\cite{Banerjee12}. Current coherence times of superconducting qubits can be as long as hundreds of microseconds~\cite{biznarova24, place21}, while we observe coupling strengths as strong as 3 MHz.  The practical requirements for observing these phenomena are well within reach, as evidenced by recent experiments demonstrating quantum simulations with systems of 10--20 qubits~\cite{karamlou24, zhang23, karamlou22}. This suggests that significant quantum effects can be studied without an exponential increase in the number of qubits, making the system scalable and practical for near-term experimental investigations.

\begin{acknowledgments}

We acknowledge Jimmy Hung and Dmytro Dubyna for help developing and maintaining the experimental setup. CMW, JHB, ZS, CXCY and IN acknowledge the Canada First Research Excellence Fund (CFREF), NSERC of Canada, the Canadian Foundation for Innovation, the Ontario Ministry of Research and Innovation, and Industry Canada for financial support. JMAC and ER acknowledge the financial support received from the IKUR Strategy under the collaboration agreement between the Ikerbasque Foundation and UPV/EHU on behalf of the Department of Education of the Basque Government. E.R. acknowledges support from the BasQ strategy of the Department of Science, Universities, and Innovation of the Basque Government. E.R. is supported by the grant PID2021-126273NB-I00 funded by MCIN/AEI/ 10.13039/501100011033 and by ``ERDF A way of making Europe" and the Basque Government through Grant No. IT1470-22. This work was supported by the EU via QuantERA project T-NiSQ grant PCI2022-132984 funded by MCIN/AEI/10.13039/501100011033 and by the European Union ``NextGenerationEU''/PRTR. This work has been financially supported by the Ministry of Economic Affairs and Digital Transformation of the Spanish Government through the QUANTUM ENIA project called – Quantum Spain project, and by the European Union through the Recovery, Transformation, and Resilience Plan – NextGenerationEU within the framework of the Digital Spain 2026 Agenda.

\end{acknowledgments}

\appendix

% \section{Appendixes}
\section{\label{app:setup} Experiment setup}
\subsection{\label{app:fridge} Fridge Setup}
\begin{figure} 
    \centering
    \includegraphics[scale =1]{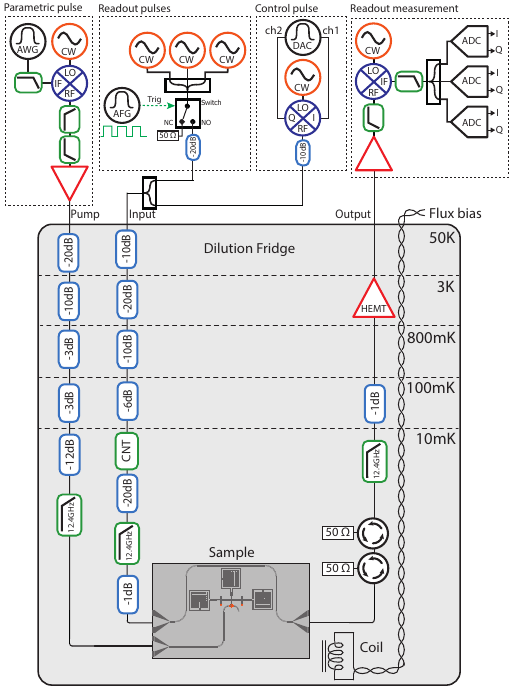}
    \caption[three-qubit Experiment Setup]{Illustration of the experimental setup.}
    \label{fig:ExperimentSetup_3Q}
\end{figure}

The device is cooled using a Bluefors dilution refrigerator that can reach a temperature of approximately 7 mK. The wiring setup, shown in Fig.~\ref{fig:ExperimentSetup_3Q}, includes three microwave lines with $50\ \Omega$ SMA cables for input, control, and output signals,  as well as a DC line terminated by a coil for external flux bias. Then, the fridge input line is heavily attenuated along the different stages to properly cool the input signal. In addition to a cryogenic low-pass filter, we use an in-house carbon nanotube-based lossy transmission line filter (CNT) designed to block thermal radiation higher than 50 GHz. At the output line, we use a low-noise amplifier at the 3K stage, namely a high electron mobility transistor (HEMT) amplifier, to enhance the signal before it reaches the fridge output for further amplification.  Also, the device output port is isolated from thermal noise using circulators at the MC stage, where any noise coming from the output line/HEMT is shunted to a $50\ \Omega$ terminated port,  thermally anchored to the MC plate. We use an external coil mounted on the sample box to apply a DC flux bias on the device via fridge DC lines. In addition, the device has an on-chip fast-flux line used for parametric control, which we connect to through the pump line. The pump is treated as the input line we discussed, heavily attenuated with a low-pass filter. 

\subsection{\label{app:measure_setup} Measurement Setup}
We use a PXDAC4800 digital-to-analog conversion (DAC) board to generate the qubit control pulses. The pulse wave is synthesized digitally at an IF frequency $f_{IF} = 150$~MHz  with a sampling rate of $1.2$ Gsps. We implement single-side-band mixing using Marki IQ mixers (MLIQ-0218) and a Rhode $\&$  Schwarz SGS continuous-wave (CW) source as an LO at around $f_{LO} = 5$ GHz. Two DAC channels are used to generate in-phase and quadrature pulses along with dc offsets calibrated to suppress the LO leakage and the lower sideband while enhancing the higher sideband at $f_{LO} + f_{IF}$.

The circuit of the readout pulses consists of three Rhode $\&$ Schwarz SGS CW sources combined and connected to the input port of a Quantic X-Microwave microwave switch.  While the normally-closed (NC) output port is terminated by a $50\ \Omega$, the normally-opened (NO) output port is connected to the fridge input.  The readout pulses are generated at the NO port by feeding a square pulse to the switch trigger port using a Tektronix arbitrary function generator (AFG). The square pulse acts as an envelope to the CW signals.

The parametric pulse we apply through the pump line is generated using an Aeroflex 3025C RF  arbitrary wave generator (AWG) with integrated mixers.  Since the AWG has an output range of only up to $6$ GHz, we employ an upconversion circuit to reach a higher frequency with filtering at the output to eliminate sidebands and LO leakages.

Finally, the fridge output is filtered and downconverted to a range of $2$ GHz before it is split into three Aeroflex 3020/30 heterodyne analog-to-digital converters (ADCs), called RF digitizers.  In the digitization process, the incoming signal is downconverted to an IF before it is sampled at $250$ Msps.  This process preserves the information from both quadratures, which are then calculated digitally by the FPGA.

Those instruments and pulses have to be synchronized together using a proper triggering scheme. All of the Aeroflex instruments are modules mounted in a NI PXI-1045 chassis with trigger routing modules on the chassis backplane.  This allows us to route the AWG trigger signal to all ADC modules.  Mounted on the same chassis, we use a NI PXI-6651 timing module to synchronize standalone instruments with the AWG trigger. It allows us to generate additional synchronized trigger signals which are connected through external cables to the AFG and DAC trigger inputs.

\section{\label{app:theory} Theory of the parametrically driven coupled transmons}
\subsection{\label{app:model} Model}
The three pairwise capacitively coupled transmon qubits in our device are described by the Hamiltonian
\begin{equation}\label{eq:ham_general}
\begin{aligned}
    \hat{\mathcal{H}} =&\sum_{jk} 4E_{\mathrm{C},jk}\hat{N}_j \hat{N}_k \\
    &-E_{\mathrm{J},1}\cos \hat{\phi}_1-E_{\mathrm{J},3}\cos \hat{\phi}_3+\hat{\mathcal{H}}_{\mathrm{sq}}.
\end{aligned}
\end{equation}
Here $E_{\mathrm{C},jj}$ and $E_{\mathrm{J},j}$ are the charging and Josephson energies of the $j$th transmon qubit, $E_{\mathrm{J},j}\gg E_{\mathrm{C},jj}$, and the operators $\hat{N}_j$ and $\hat{\phi}_j$ are the     Cooper pair number and superconducting phase difference and are conjugate to each other. The off-diagonal term $E_{\mathrm{C},jk}$ with $j\neq k$ describes the capacitive coupling between transmons $j$ and $k$; generally $|E_{\mathrm{C},jk}|\ll E_{\mathrm{C},jj}$. The charging energy matrix $E_{\mathrm{C}}$ is related to the qubit capacitances $C_{j}$ and the coupling capacitances $C_{jk}$ via
% , where $\hat{\phi}_2\equiv \hat{\phi}_{\mathrm{sq}}$
%
\begin{widetext}
\begin{equation}
    E_{\mathrm{C}}=\frac{e^2}{2}\begin{pmatrix}
    C_1+C_{12}+C_{13} & -C_{12} & -C_{13}\\
    -C_{12} & C_2+C_{12}+C_{23} & -C_{23}\\
    -C_{13} & -C_{23} & C_3+C_{13}+C_{23}
    \end{pmatrix}^{-1}.
\end{equation}
\end{widetext}
The tunable qubit 2 contains a SQUID consisting of two Josephson junctions with Josephson energies $E_{\mathrm{J},L}$ and $E_{\mathrm{J},R}$:

\begin{equation}\label{eq:ham_sq}
\begin{aligned}
    \hat{\mathcal{H}}_{\mathrm{sq}}=&-E_{\mathrm{J},L}\cos (\hat{\phi}_{2}-\frac{\pi\hat{\Phi}_\mathrm{ext}}{\Phi_0}) \\
    &-E_{\mathrm{J},R}\cos (\hat{\phi}_{2}+\frac{\pi\hat{\Phi}_\mathrm{ext}}{\Phi_0})\\
    =& -E_{\mathrm{J}} \cos\frac{\pi\hat{\Phi}_{\mathrm{ext}}}{\Phi_0}\cos \hat{\phi}_{2}\\
    &-\delta E_\mathrm{J} \sin \frac{\pi\hat{\Phi}_{\mathrm{ext}}}{\Phi_0}\sin \hat{\phi}_{2}.
\end{aligned}
\end{equation}
%
%\begin{align} \hat{\mathcal{H}}(t)&=\hat{\mathcal{H}}_1+\hat{\mathcal{H}}_2 (t)+\hat{\mathcal{H}}_3+\hat{\mathcal{H}}_\mathrm{c},\\ \hat{\mathcal{H}}_j&=4E_{\mathrm{C},jj}\hat{N}_j^2-E_{\mathrm{J},j}\cos \hat{\phi}_j\, (j=1,3),\\ \hat{\mathcal{H}}_2 (t)&=4E_{\mathrm{C},2}\hat{N}_2^2-E_{\mathrm{J},L}\cos (\hat{\phi}_{\mathrm{sq}}-\frac{\pi\Phi_\mathrm{ext}(t)}{\Phi_0})-E_{\mathrm{J},R}\cos (\hat{\phi}_{\mathrm{sq}}+\frac{\pi\Phi_\mathrm{ext}(t)}{\Phi_0}),\\ \hat{\mathcal{H}}_\mathrm{c} &= \sum_{j<k} 8E_{\mathrm{C},jk}\hat{N}_j \hat{N}_k. \end{align}
%
Here $E_{\mathrm{J}}= E_{\mathrm{J},L}+E_{\mathrm{J},R}$, and $\delta E_\mathrm{J}=E_{\mathrm{J},L}-E_{\mathrm{J},R}$. The SQUID is threaded by an external flux $\Phi_\mathrm{ext}(t)=\Phi_{p}(t) + \Phi_{b} $ which consists of a dc flux bias $\Phi_{b}$ and an ac parametric drive $\Phi_{p}(t)$. We work at zero bias $\Phi_b = 0$ and apply the parametric approximation to represent $\Phi_p(t)$ as a classical signal $\alpha_p(t)$. This allows us to express the Hamiltonian as $\hat{\mathcal{H}}=\hat{\mathcal{H}}_0+\hat{\mathcal{H}}'(t)$, where the time-independent part is
\begin{equation}\label{eq:h0_nphi}
\begin{aligned}
    \hat{\mathcal{H}}_0=&\sum_{jk} 4E_{\mathrm{C},jk}\hat{N}_j \hat{N}_k\\
    &-\sum_{j=1,3} E_{\mathrm{J},j}\cos \hat{\phi}_j-E_{\mathrm{J}}\cos \hat{\phi}_2,
\end{aligned}
\end{equation}
and the drive part reads
\begin{equation}\label{eq:hpr_drive}
\begin{aligned}
    \hat{\mathcal{H}}'(t)=&-E_{\mathrm{J}} (\cos\frac{\pi\alpha_{p}(t)}{\Phi_0}-1)\cos \hat{\phi}_{2}\\
    &-\delta E_\mathrm{J} \sin \frac{\pi\alpha_{p}(t)}{\Phi_0}\sin \hat{\phi}_{2}.
\end{aligned}
\end{equation}
%
%It is straightforward to rewrite Eq.~\eqref{eq:ham_sq_0} as
%\begin{equation}\label{eq:ham_sq}\hat{\mathcal{H}}_{\mathrm{sq}} ,\end{equation}

\subsection{\label{app:3qi} Three-qubit interaction}
We now demonstrate analytically how to activate the three-qubit interaction. We first expand the cosine terms in $\hat{\mathcal{H}}_0$ to the quartic order and rewrite $\hat{\mathcal{H}}_0$ in terms of creation and annihilation operators. In the rotating wave approximation, discarding all terms that do not conserve the number of excitations, we have
\begin{equation}
\begin{aligned}\label{eq:h0_boson}
    \hat{\mathcal{H}}_0\approx&\sum_{j} [\hbar\omega_{0, j}\hat{b}_j^{\dagger} \hat{b}_j-\frac{E_{\mathrm{C}, jj}}{2}\hat{b}_j^{\dagger}\hat{b}_j^{\dagger}\hat{b}_j \hat{b}_j]\\
    &+ \sum_{j< k} g_{jk}(\hat{b}_j^{\dagger} \hat{b}_k+\mathrm{h.c.}),
    %\hat{\mathcal{H}}_0\approx\sum_{j} [\hbar\omega_{0, j}\hat{b}_j^{\dagger} \hat{b}_j-\frac{1}{12}E_{\mathrm{C}, jj}(\hat{b}_j^{\dagger}+\hat{b}_j)^4] - \sum_{j\neq k} g_{jk}(\hat{b}_j^{\dagger}-\hat{b}_j)(\hat{b}_k^{\dagger}-\hat{b}_k),
\end{aligned}
\end{equation}
where $\omega_{0, j}$ are the bare qubit frequencies,
\begin{equation}
    \hbar\omega_{0, j} = \sqrt{8E_{\mathrm{C}, jj} E_{\mathrm{J}, j}}-E_{\mathrm{C}, jj},
\end{equation}
the capacitive coupling constants are
\begin{equation}\label{eq:g_jk_capacitive}
    g_{jk}=2E_{\mathrm{C}, jk}\left(\frac{E_{\mathrm{J}, j}E_{\mathrm{J}, k}}{4E_{\mathrm{C}, jj}E_{\mathrm{C}, kk}}\right)^{\frac{1}{4}},
\end{equation}
and the creation/annihilation operators are given by
\begin{equation}
\begin{aligned}
    \hat{\phi}_j =& \left(\frac{2E_{\mathrm{C}, jj}}{E_{\mathrm{J}, j}}\right)^{\frac{1}{4}}(\hat{b}_j^{\dagger}+\hat{b}_j),\\
    \hat{N}_j =& \frac{i}{2}\left(\frac{E_{\mathrm{J}, j}}{2E_{\mathrm{C}, jj}}\right)^{\frac{1}{4}}(\hat{b}_j^{\dagger}-\hat{b}_j),
     %\hat{b}_j=\left(\frac{E_{\mathrm{J}, j}}{2E_{\mathrm{C}, jj}}\right)^{\frac{1}{4}}\frac{\hat{\phi}_j}{2}-i\left(\frac{2E_{\mathrm{C}, jj}}{E_{\mathrm{J}, j}}\right)^{\frac{1}{4}}\hat{N}_j.
\end{aligned}
\end{equation}
and $E_{\mathrm{J},2}=E_{\mathrm{J}}$. To obtain the frequencies of the coupled qubits $\omega_{j}$, we diagonalize the quadratic part of Eq.~\eqref{eq:h0_boson}, $\hat{b}_j=\sum_k U_{jk}\hat{a}_k$, such that
\begin{equation}\label{eq:h0_boson_a}
\begin{aligned}
    \hat{\mathcal{H}}_0\approx&\sum_{j}(\hbar\omega_{j}\hat{a}_j^{\dagger} \hat{a}_j+\frac{\hbar K_{j}}{2}\hat{a}_{j}^{\dagger}\hat{a}_{j}^{\dagger}\hat{a}_{j}\hat{a}_{j})\\
    &+\sum_{j< k}\hbar \chi_{jk}\hat{a}_{j}^{\dagger}\hat{a}_{j}\hat{a}_{k}^{\dagger}\hat{a}_{k}+\dots 
    %+O(\hat{a}^{\dagger}\hat{a}^{\dagger}\hat{a}\hat{a}),
    %
\end{aligned}
\end{equation}
where $K_{j}$ and $\chi_{jk}$ are self- and cross-Kerr nonlinearities respectively, and for simplicity we ignore other terms in the interaction which mix the 2-photon states. Importantly, the unitary matrix $U_{jk}$ is generally not diagonal because of the nonzero $g_{jk}$.

We can now expand the drive Hamiltonian Eq.~\eqref{eq:hpr_drive} in powers of $\hat{a}_{j}$,
\begin{equation}\label{eq:hpr_drive_a}
    \hat{\mathcal{H}}'(t)=-\sum_n g_n(\alpha_p (t)) [\sum_{k=1}^3 (U^*_{2k}\hat{a}_k^\dagger +U_{2k}\hat{a}_k)]^n,
\end{equation}
where the coefficients $g_n$ are functions of $\alpha_p$. In the interaction picture where the time evolution of operators is governed by $\hat{\mathcal{H}}_0$, different terms in Eq.~\eqref{eq:hpr_drive_a} have distinct time dependence. A term is slow-varying and does not vanish in the rotating wave approximation only when $\alpha_p (t)$ has a frequency component that matches the time dependence of that term. In particular, the three-body interaction term $\hat{a}_1^{\dagger}\hat{a}_2^{\dagger}\hat{a}_3$ has the frequency
\begin{equation}\label{eq:three_body_freq}
    \omega_{001 \rightarrow 110} = (\omega_1 + \omega_2 + \chi_{12}) - \omega_3;
\end{equation}
therefore, when we drive $\alpha_p (t)$ at the frequency $\omega_{001 \rightarrow 110}$, $\alpha_p(t)=A_p\cos(\omega_{001 \rightarrow 110} t+\phi)$, we obtain a term in the Hamiltonian
\begin{equation}\label{eq:h_int_3q_sm}
    \hat{\mathcal{H}}_{\mathrm{int}}\approx -J(A_p) \hat{a}_1^{\dagger}\hat{a}_2^{\dagger}\hat{a}_3 +\mathrm{h.c.}.
\end{equation}
This term originates from the cubic term in the expansion Eq.~\eqref{eq:hpr_drive_a} which is traced back to the sine term in Eq.~\eqref{eq:hpr_drive}. Therefore, the coefficient $J(A_p)\propto \delta E_{\mathrm{J}}A_p$ to the leading order in $A_p$ [see Eq.~\eqref{eq:pert_g_eff}].

\subsection{\label{app:numerics} Rabi oscillations and perturbation theory}
In our numerical simulations of the spectrum and the dynamics of the system, we employ the scQubits~\cite{scqubit2021,scqubit2022} and QuTiP~\cite{qutip} libraries, working with the Hamiltonians Eqs.~\eqref{eq:h0_nphi} and \eqref{eq:hpr_drive} without assuming weak anharmonicity and the rotating wave approximation. We find the spectrum $\{\epsilon_l \}$ of the coupled transmon system Eq.~\eqref{eq:h0_nphi} in the eigenbases of the individual transmons. The dynamics due to the drive term Eq.~\eqref{eq:hpr_drive} are then simulated in the eigenbasis $\ket{l}$ of Eq.~\eqref{eq:h0_nphi}.
%\begin{equation}
%    \hat{\mathcal{H}}_0=\sum_{l=1}^{L_{\mathrm{cutoff}}}|l \rangle\langle l|, \hat{\mathcal{H}}'(t)=\sum_{ll'} |l \rangle\langle l|\hat{\mathcal{H}}'(t)|l'\rangle\langle l'|.
%\end{equation}

For weak parametric drives, we can calculate both the three-body interaction strength and the shift of the resonance frequency perturbatively. To the first order in $A_p$, the three-body interaction strength is simply
\begin{equation}\label{eq:pert_g_eff}
    J(A_p)=\frac{\pi\delta E_\mathrm{J} }{2\Phi_0}\langle 110|\sin \hat{\phi}_{2}| 001 \rangle A_p,
\end{equation}
where the states $\ket{110}$ and $\ket{001}$ are eigenstates of coupled transmon system Eq.~\eqref{eq:h0_nphi}; in particular we identify $\ket{110}$ by the method in Sec.~\ref{sec:characterize}.

On the other hand, to the second order in $A_p$, there are two distinct contributions to the shift of the resonance frequency Eq.~\eqref{eq:three_body_freq}. The diagonal matrix elements of the cosine term in Eq.~\eqref{eq:hpr_drive} (i.e. the static qubit dispersion) contribute to the shift in first-order perturbation theory. After averaging over time, these diagonal matrix elements have the form:
\begin{equation}
\begin{aligned}
    \overline{\langle l|\hat{\mathcal{H}}'(t)|l\rangle } &\approx\frac{\pi^2}{2\Phi_0^2} E_{\mathrm{J}} \overline{\alpha_{p}(t)^2}\langle l|\cos \hat{\phi}_{2}|l\rangle\\
    &=\frac{\pi^2 A_p^2}{4\Phi_0^2} E_{\mathrm{J}}\langle l|\cos \hat{\phi}_{2}|l\rangle,
\end{aligned}
\end{equation}
where the overline denotes the time average over many periods. Meanwhile, the sine term in Eq.~\eqref{eq:hpr_drive} leads to a Bloch--Siegert-type shift that is accounted for in second-order perturbation theory~\cite{Shirley1965periodic}. The total second-order shift of the resonance frequency reads
\begin{equation}
\begin{aligned}\label{eq:pert_delta_omega_3q}
    &\omega_{3q}-\omega_{3q}^{(0)} = \frac{\pi^2 A_p^2}{4\Phi_0^2} \Bigg\{\delta E_{\mathrm{J}}^2 \sum_{l}\Big[\frac{|\langle l|\sin \hat{\phi}_{2}|110\rangle|^2}{\epsilon_{110}-(\epsilon_l-\omega_{3q}^{(0)})}\\
    &+\frac{|\langle l|\sin \hat{\phi}_{2}|110\rangle|^2}{\epsilon_{110}-(\epsilon_{l}+\omega_{3q}^{(0)})} -\frac{|\langle l|\sin \hat{\phi}_{2}|001\rangle|^2}{\epsilon_{001}-(\epsilon_l-\omega_{3q}^{(0)})}\\
    &-\frac{|\langle l|\sin \hat{\phi}_{2}|001\rangle|^2}{\epsilon_{001}-(\epsilon_{l}+\omega_{3q}^{(0)})}\Big]+E_{\mathrm{J}}(\langle 110|\cos \hat{\phi}_{2}|110\rangle\\
    &-\langle 001|\cos \hat{\phi}_{2}|001\rangle) \Bigg\},
\end{aligned}
\end{equation}
where $\omega_{3q}^{(0)}=\omega_{001 \rightarrow 110}$. The perturbative results for the three-body interaction strength and the resonance frequency, Eqs.~\eqref{eq:pert_g_eff} and \eqref{eq:pert_delta_omega_3q}, are shown in Fig.~\ref{fig:ELevels}(c), together with experimental data and brute-force time evolution results obtained using the fit parameters. We emphasize that the $\cos \hat{\phi}_{2}$ term in Eq.~\eqref{eq:pert_delta_omega_3q} alone does not yield the correct perturbative shift for our strongly coupled system.

\section{\label{app:map}Mapping the transmon chain to LGT}
\begin{figure}[h]
    \centering
    \includegraphics[scale=0.16]{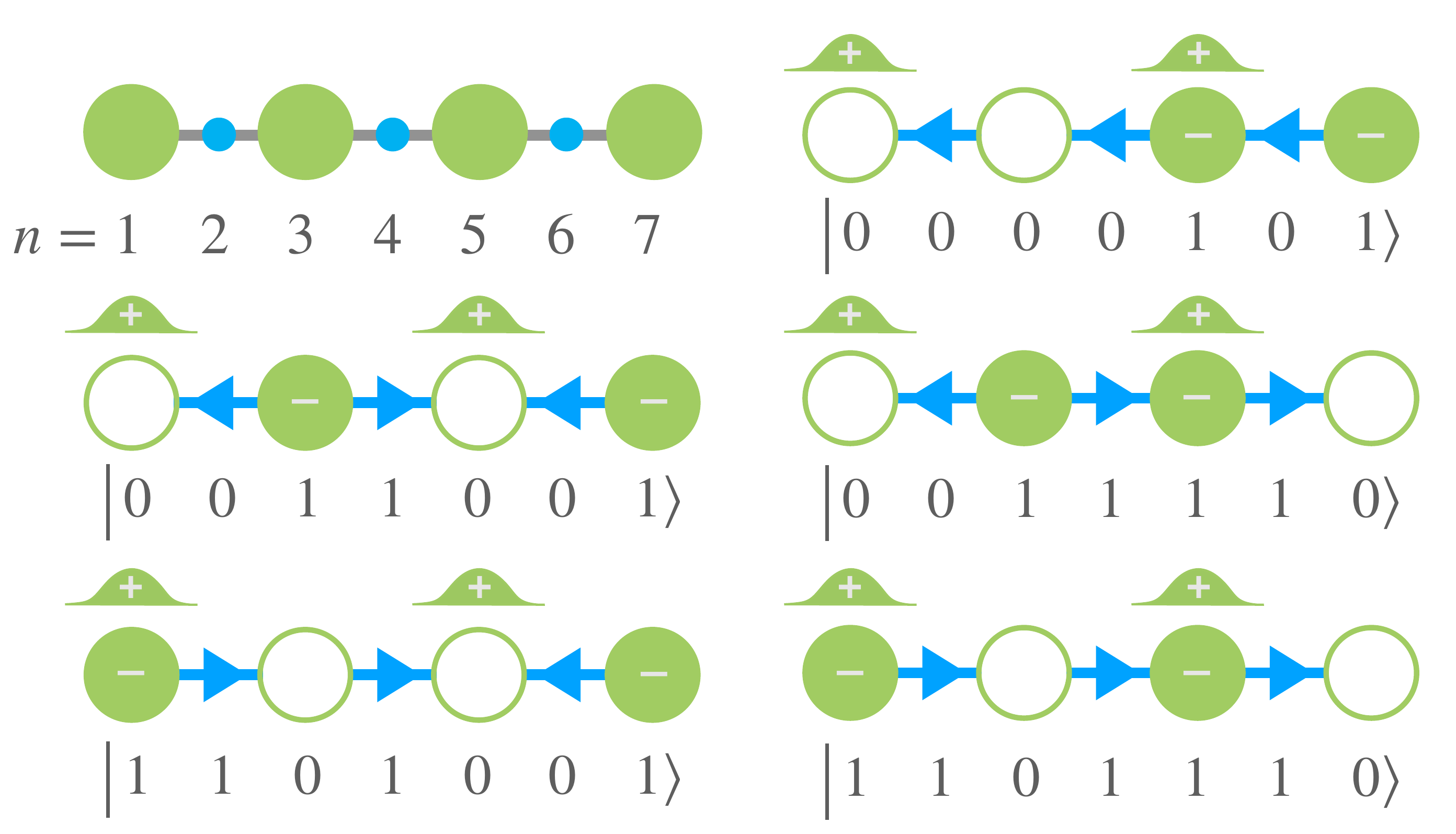}
    \caption{The 7-transmon chain with open boundary conditions and its 5 states allowed by Gauss's law.} 
    \label{fig:7transmonstates}
\end{figure}

To illustrate the mapping of the transmon Hamiltonian to the $U(1)$ LGT in a system with more than three transmons, we consider a 7-transmon chain with open boundaries (see Fig.~\ref{fig:7transmonstates}), where four transmons serve as matter sites. Adopting the two-level approximation and choosing the gauge conditions $\hat{G}_{n}\left|\psi\right\rangle =0$ for $1\leq n\leq4$, we find five states allowed by Gauss's law:
\begin{equation}
\begin{aligned}
\left|\mathrm{I}\right\rangle=&\left|0000101\right\rangle ,\\
E_{\mathrm{I}}/\hbar  =&\omega_{5}+\omega_{7},\\
\left|\mathrm{II}\right\rangle=&\left|0011001\right\rangle ,\\
E_{\mathrm{II}}/\hbar =&\omega_{3}+\omega_{4}+\omega_{7}+\chi_{34},\\
\left|\mathrm{III}\right\rangle=&\left|0011110\right\rangle ,\\
E_{\mathrm{III}}/\hbar  =&\omega_{3}+\omega_{4}+\omega_{5}+\omega_{6}+\chi_{34}+\chi_{45}+\chi_{56},\\
\left|\mathrm{IV}\right\rangle=&\left|1101001\right\rangle ,\\
E_{\mathrm{IV}}/\hbar =&\omega_{1}+\omega_{2}+\omega_{4}+\omega_{7}+\chi_{12},\\
\left|\mathrm{V}\right\rangle=&\left|1101110\right\rangle ,\\
E_{\mathrm{V}}/\hbar  =&\omega_{1}+\omega_{2}+\omega_{4}+\omega_{5}+\omega_{6}\\
&+\chi_{12}+\chi_{45}+\chi_{56}.
\end{aligned}
\end{equation}
%
%\begin{comment}
%
%
%\end{comment}
%
\begin{comment}
\begin{figure}[h]
    \centering
    \includegraphics[scale=0.7]{Additional_figures/seventransmons_2.pdf}
    \caption{The 7-transmon chain with open boundary conditions and its 5 states allowed by Gauss's law.} 
    \label{fig:7transmonstates}
\end{figure}
\end{comment}
%
These five states are coupled by the three-body term $\hat{\sigma}_{n}^{+}\hat{\tau}_{n,n+1}^{+}\hat{\sigma}_{n+1}^{-}$ ($1\leq n \leq 3$). In particular, $n=1$ couples $\left|\mathrm{II}\right\rangle $ to $\left|\mathrm{IV}\right\rangle $ and $\left|\mathrm{III}\right\rangle $ to $\left|\mathrm{V}\right\rangle $, $n=2$ couples $\left|\mathrm{I}\right\rangle $ to $\left|\mathrm{II}\right\rangle $, and $n=3$ couples $\left|\mathrm{II}\right\rangle $ to $\left|\mathrm{III}\right\rangle $ and $\left|\mathrm{IV}\right\rangle $ to $\left|\mathrm{V}\right\rangle $. Therefore, in the rotating wave approximation, the Hamiltonian in the gauge-invariant sector reads
\begin{equation}
\begin{aligned}
&\hat{\mathcal{H}} =\sum_{n=\mathrm{I}}^{\mathrm{V}}E_{n}\left|n\right\rangle \left\langle n\right|\\
&-\Bigg[J_{1}e^{-i\left(\omega_{3q,1}+\delta\omega_{1}\right)t}\left(\left|\mathrm{II}\right\rangle \left\langle \mathrm{IV}\right|+\left|\mathrm{III}\right\rangle \left\langle \mathrm{V}\right|\right)\\
 & +J_{2}e^{-i\left(\omega_{3q,2}+\delta\omega_{2}\right)t}\left|\mathrm{I}\right\rangle \left\langle \mathrm{II}\right|\\
 & +J_{3}e^{-i\left(\omega_{3q,3}+\delta\omega_{3}\right)t}\left(\left|\mathrm{II}\right\rangle \left\langle \mathrm{III}\right|+\left|\mathrm{IV}\right\rangle \left\langle \mathrm{V}\right|\right)+\text{h.c.}\Bigg],
\end{aligned}
\end{equation}
where the associated three-body resonant frequencies $\omega_{3q,n}$ at the zeroth order in three-body interaction strengths (for simplicity we suppress the superscript $(0)$) are
\begin{equation}\label{eq:omega_3q_7transmon}
\begin{aligned}
\omega_{3q,1} & =(E_{\mathrm{IV}}-E_{\mathrm{II}})/\hbar=(E_{\mathrm{V}}-E_{\mathrm{III}})/\hbar\\
&=\omega_{1}+\omega_{2}-\omega_{3}+\chi_{12}-\chi_{34},\\
\omega_{3q,2} & =(E_{\mathrm{II}}-E_{\mathrm{I}})/\hbar=\omega_{3}+\omega_{4}-\omega_{5}+\chi_{34},\\
\omega_{3q,3} & =(E_{\mathrm{III}}-E_{\mathrm{II}})/\hbar=(E_{\mathrm{V}}-E_{\mathrm{IV}})/\hbar\\
&=\omega_{5}+\omega_{6}-\omega_{7}+\chi_{45}+\chi_{56},
\end{aligned}
\end{equation}
and $\delta\omega_{n}$ are the pump detunings from these frequencies.

We now go to the interaction picture to remove the time dependence of the Hamiltonian and map it to the $U(1)$ LGT. Denoting the reference energies as $\epsilon_{n}$, we arrive at the following effective Hamiltonian:
\begin{equation}\label{eq:H_rot_7transmon}
\begin{aligned}
\hat{\mathcal{H}}=&\sum_{n=\mathrm{I}}^{\mathrm{V}}\left(E_{n}-\epsilon_{n}\right)\left|n\right\rangle \left\langle n\right|\\
-&\Bigg[J_{1}e^{-i[\omega_{3q,1}+\delta\omega_{1}-(\epsilon_{\mathrm{IV}}-\epsilon_{\mathrm{II}})/\hbar]t}\left|\mathrm{II}\right\rangle \left\langle \mathrm{IV}\right|
+\cdots\Bigg].
\end{aligned}
\end{equation}
This should be compared with the target Hamiltonian to be simulated:
\begin{equation}\label{eq:H_target_7transmon}
\begin{aligned}
\hat{\mathcal{H}} =& \frac{1}{2} \sum_{n=1}^{4} (-1)^n \mu_n\hat{\sigma}^z_n \\
& - \sum_{n=1}^{3} J_n(\hat{\sigma}_n^{+} \hat{\tau}^+_{n,n+1} \hat{\sigma}^-_{n+1} + \mathrm{h.c.}). 
\end{aligned}
\end{equation}
To eliminate the time dependence in Eq.~\eqref{eq:H_rot_7transmon}, we require
\begin{equation}\label{eq:condition_time}
\begin{aligned}
\hbar(\omega_{3q,1}+\delta\omega_{1}) & =\epsilon_{\mathrm{IV}}-\epsilon_{\mathrm{II}}=\epsilon_{\mathrm{V}}-\epsilon_{\mathrm{III}},\\
\hbar(\omega_{3q,2}+\delta\omega_{2}) & =\epsilon_{\mathrm{II}}-\epsilon_{\mathrm{I}},\\
\hbar(\omega_{3q,3}+\delta\omega_{3}) & =\epsilon_{\mathrm{III}}-\epsilon_{\mathrm{II}}=\epsilon_{\mathrm{V}}-\epsilon_{\mathrm{IV}}.
\end{aligned}
\end{equation}
In addition, to reproduce the diagonal elements of the target Hamiltonian Eq.~\eqref{eq:H_target_7transmon}, we should have
\begin{equation}\label{eq:condition_diag}
\begin{aligned}
E_{\mathrm{I}}-\epsilon_{\mathrm{I}} & =\frac{1}{2}(-\mu_{1}+\mu_{2}+\mu_{3}-\mu_{4}),\\
E_{\mathrm{II}}-\epsilon_{\mathrm{II}} & =\frac{1}{2}(-\mu_{1}-\mu_{2}-\mu_{3}-\mu_{4}),\\
E_{\mathrm{III}}-\epsilon_{\mathrm{III}} & =\frac{1}{2}(-\mu_{1}-\mu_{2}+\mu_{3}+\mu_{4}),\\
E_{\mathrm{IV}}-\epsilon_{\mathrm{IV}} & =\frac{1}{2}(\mu_{1}+\mu_{2}-\mu_{3}-\mu_{4}),\\
E_{\mathrm{V}}-\epsilon_{\mathrm{V}} & =\frac{1}{2}(\mu_{1}+\mu_{2}+\mu_{3}+\mu_{4}).
\end{aligned}
\end{equation}
Subtracting Eq.~\eqref{eq:omega_3q_7transmon} from Eq.~\eqref{eq:condition_time} and making use of Eq.~\eqref{eq:condition_diag}, we find
\begin{equation}\label{eq:condition_mu}
\begin{aligned}
\hbar\delta\omega_{1} & =-\mu_{1}-\mu_{2},\\
\hbar\delta\omega_{2} & =\mu_{2}+\mu_{3},\\
\hbar\delta\omega_{3} & =-\mu_{3}-\mu_{4}.
\end{aligned}
\end{equation}
Equation~\eqref{eq:condition_mu} shows that we can adjust three out of the four fermion mass terms independently by changing the detuning of the three-body parametric drives. Similar relations can be derived for longer chains.

\section{\label{app:characterize} Device Characterization}
The resonators are capacitively coupled to a two-port (through) transmission line. We identify the resonators in our device using $S_{21}$ VNA measurements. The measured parameters of the readout resonators are listed in Table~\ref{tab:Resonators}. We note that resonator 3 was far off from design with a very large internal decay rate, $\kappa_{\mathrm{int}}$, which posed challenges on the qubit state readout.
\begin{table} 
\centering
\setlength{\tabcolsep}{10pt}
\begin{tabular}{|l|c|c|c|}
\hline
              \textbf{Readout resonator }             & 1 & 2 & 3\\
\hline
\textbf{ $\omega_r/2\pi$ [GHz] }        & $7.698$       &  $7.518$      &  $7.035$        \\
\textbf{ $\kappa_{\mathrm{int}}/2\pi$ [MHz]}    & $0.439$             & $0.489$             & $5.1$                 \\
\textbf{ $\kappa/2\pi$ [MHz]}              & $0.650$              & $0.643$            & $6.37$                 \\
\textbf{ $\eta$}    & $0.325$               & $0.240$               & $0.199$              \\
\hline 
\end{tabular} 
\caption[Readout Resonators characteristics]{ Resonator parameters extracted from VNA measurements, including resonance frequencies, internal and external decay rates.}
\label{tab:Resonators}
\end{table}

With one-tone and two-tone spectroscopy, we identify the qubit frequencies and the higher energy levels to calibrate the higher-order transitions required for the three-body interaction. We measure the qubit frequencies as a function of the flux bias, $\phi_b$. Finally, we perform qubits characterization which are listed in Table~\ref{tab:qubit_characterize}.
\begin{table}
\centering
\setlength{\tabcolsep}{10pt}
\begin{tabular}{|l|c|c|c|}
\hline
Qubit                           & \textbf{ 1} & \textbf{ 2} & \textbf{ 3} \\
\hline
\textbf{ $\omega_q/2\pi$ [GHz]} & $5.725$            & $5.910$            & $5.055$             \\
% \textbf{Qubit Anharmonicity [MHz]}       & $-184$             & $-105$             & $-156$              \\
\textbf{ $T_1$ [ns]}                         & $4216$& $1302$& $4152 \pm 2426$\\
\textbf{ $T_{\mathrm{Ramsey}}$ [ns]}                    & $2470$& $971$& $2907$              \\
\textbf{ $T_{\mathrm{Echo}}$ [ns]}                      & -& $965$& $4527$              \\
\textbf{ $2\chi^{R1}/2\pi$ [MHz]}    & $-7.3$             & $-0.4$             & $0$                 \\
\textbf{ $2\chi^{R2}/2\pi$ [MHz]}    & $-2.2$             & $-2.4$             & $0$                 \\
\textbf{ $2\chi^{R3}/2\pi$ [MHz]}    & $-2$               & $-3$               & $-0.5$              \\
\hline 
\end{tabular}
\caption[three-qubit device characterization]{Qubit frequencies, anharmonicities, lifetimes, and dispersive shifts of resonator frequencies extracted during device characterization.}
    \label{tab:qubit_characterize}
\end{table}

\section{\label{app:extract} Extracting the qubit state from resonators}
To characterize the three-body interaction, we initialize the system in the $\ket{001}$ state and let the system evolve under the three-body interaction for an evolution time $t_{\mathrm{evolve}}$. We then apply a readout pulse to each resonator and measure the resonator response as a function of the measurement time $t_{\mathrm{meas}}$.%, and integrate over a $t_{\mathrm{meas}}$ window.

In the following, we describe how to fit the measurement results. We find the resonator response is approximately independent of how many multiplexed readout pulses are applied simultaneously. Therefore, only one resonator is considered at a time for simplicity. Since coherent evolution only takes place in the subspace spanned by $\ket{110}$ and $\ket{001}$, it is enough to consider an effective Hamiltonian with five qubit states, $s=\ket{001}, \ket{110}, \ket{100}, \ket{010}$ and $\ket{000}$, and one resonator $\hat{a}$:
%
%\begin{align}
%\frac{\hat{H}_{\text{cB}}}{\hbar}=&\left(\omega_{r}+2\sum_{s}\chi_{s}\left|s\right\rangle \left\langle s\right|\right)\hat{a}^{\dagger}\hat{a}+\sum_{s}\omega_{s}\left|s\right\rangle \left\langle s\right|+\left(\epsilon_{m}\left(t\right)\hat{a}^{\dagger}e^{-i\omega_{m}t}+\epsilon_{m}^{*}\left(t\right)\hat{a}e^{i\omega_{m}t}\right)\nonumber\\&
%+\left(\Omega\left(t\right)\ket{k} \bra{l}e^{-i\omega_{p}t}+\Omega^{*}\left(t\right)\ket{l} \bra{k}e^{i\omega_{p}t}\right),
%\end{align}
%
\begin{equation}
\begin{aligned}
\hat{H}_{\text{cB}}/\hbar =& (\omega_{r}+2\sum_{s}\chi_{s}\left|s\right\rangle \left\langle s\right|) \hat{a}^{\dagger} \hat{a} \\ &+\alpha \hat{a}^{\dagger} \hat{a}^{\dagger} \hat{a} \hat{a} + \sum_{s} \omega_{s} \left|s\right\rangle \left\langle s\right| \\
&+\left(\epsilon_{m}\left(t\right)\hat{a}^{\dagger}e^{-i\omega_{m}t}+\epsilon_{m}^{*}\left(t\right)\hat{a}e^{i\omega_{m}t}\right)\\
&+(\Omega\left(t\right)\ket{110} \bra{001} e^{-i\omega_{p}t}\\
& +\Omega^{*}\left(t\right)\ket{001} \bra{110}e^{i\omega_{p}t} ),
\end{aligned}
\end{equation}
where $\omega_r$ and $\alpha$ are the frequency and the nonlinearity of the resonator mode $\hat{a}$, $\omega_s$ and $2\chi_s$ are the frequency and the dispersive shift for the qubit state $s$ respectively, $\Omega(t)$ is the three-body interaction pulse at frequency $\omega_p \approx \omega_{001 \rightarrow 110}$ lasting for $t_{\mathrm{evolve}}$, and $\epsilon_m(t)$ is the readout pulse at frequency $\omega_m$ turned on after $t_{\mathrm{evolve}}$. According to the input-output theory, the output signal is then given by $\epsilon_m(t)-(i/2)\kappa_{\mathrm{ext}}\left\langle \hat{a}(t)\right\rangle$, where $\kappa_{\mathrm{ext}}$ is again the external coupling rate of the resonator to the transmission line, and the factor of $1/2$ comes from measuring the resonator in the hanger mode~\cite{gao08}.

Following Ref.~\cite{Bianchetti09}, we solve a set of differential equations for the expectation values of qubit operators and the resonator field $\hat{a}$. These equations are truncated by factoring the higher-order terms, e.g. $\langle \hat{a}^\dagger \hat{a} \ket{001}\bra{110}\rangle \approx \langle \hat{a}^\dagger \hat{a} \rangle \langle \ket{001}\bra{110} \rangle$ and $\langle \hat{a}^\dagger \hat{a} \hat{a} \ket{001}\bra{110}\rangle \approx \langle \hat{a}^\dagger \hat{a} \rangle \langle \hat{a} \ket{001}\bra{110} \rangle$. In the rotating frame, the equations read
\begin{subequations}
    \begin{equation}
        \begin{aligned}
            &\frac{d}{dt}\left\langle \ket{001}\bra{001}\right\rangle \\
            =&-i\left[\Omega^{*}\left(t\right)\left\langle \ket{001}\bra{110}\right\rangle -\Omega\left(t\right)\left\langle \ket{110}\bra{001}\right\rangle \right]\\
            -&\gamma_{001\to000}\left\langle \ket{001}\bra{001}\right\rangle +\gamma_{000\to001}\left\langle \ket{000}\bra{000}\right\rangle,
        \end{aligned}
    \end{equation}
    \begin{equation}
        \begin{aligned}
            &\frac{d}{dt}\left\langle \ket{110}\bra{110}\right\rangle\\
            =&i\left[\Omega^{*}\left(t\right)\left\langle \ket{001}\bra{110}\right\rangle -\Omega\left(t\right)\left\langle \ket{110}\bra{001}\right\rangle \right]\\
            -&\left(\gamma_{110\to100}+\gamma_{110\to010}\right)\left\langle \ket{110}\bra{110}\right\rangle \\
            +&\gamma_{100\to110}\left\langle \ket{100}\bra{100} \right\rangle + \gamma_{010\to110} \left\langle \ket{010}\bra{010}\right\rangle,
        \end{aligned}
    \end{equation}

    \begin{equation}
        \begin{aligned}
            &\frac{d}{dt}\left\langle \ket{100}\bra{100}\right\rangle \\
            =&-\left(\gamma_{100\to000}+\gamma_{100\to110}\right)\left\langle \ket{100}\bra{100}\right\rangle \\
            +&\gamma_{110\to100}\left\langle \ket{110}\bra{110}\right\rangle +\gamma_{000\to100}\left\langle \ket{000}\bra{000}\right\rangle,
        \end{aligned}
    \end{equation}

    \begin{equation}
        \begin{aligned}
            &\frac{d}{dt}\left\langle \ket{010}\bra{010}\right\rangle \\
            =&-\left(\gamma_{010\to000}+\gamma_{010\to110}\right)\left\langle \ket{010}\bra{010}\right\rangle \\
            +&\gamma_{110\to010}\left\langle \ket{110}\bra{110}\right\rangle +\gamma_{000\to010}\left\langle \ket{000}\bra{000}\right\rangle,
        \end{aligned}
    \end{equation}

    \begin{equation}
        \begin{aligned}
            &\frac{d}{dt}\left\langle \ket{000}\bra{000}\right\rangle\\ 
            =&\gamma_{100\to000}\left\langle \ket{100}\bra{100}\right\rangle +\gamma_{010\to000}\left\langle \ket{010}\bra{010}\right\rangle \\
            +&\gamma_{001\to000}\left\langle \ket{001}\bra{001}\right\rangle -(\gamma_{000\to100}+\gamma_{000\to010}\\
            +&\gamma_{000\to001})\times\left\langle \ket{000}\bra{000}\right\rangle,
        \end{aligned}
    \end{equation}

    \begin{equation}
        \begin{aligned}
            &\frac{d}{dt}\left\langle \ket{110}\bra{001}\right\rangle\\ 
            =&i[\omega_{110}-\omega_{001}-\omega_{p}+2\left(\chi_{110}-\chi_{001}\right)\left\langle\hat{a}^{\dagger}\hat{a}\right\rangle]\\
            &\times\left\langle \ket{110}\bra{001}\right\rangle \\
            +&i\Omega^{*}\left(t\right)\left[\left\langle \ket{001}\bra{001}\right\rangle -\left\langle \ket{110}\bra{110}\right\rangle \right]\\
            -&[\frac{1}{2}\left(\gamma_{110\to100}+\gamma_{110\to010}+\gamma_{001\to000}\right)+\gamma_{\phi}]\\
            &\times\left\langle \ket{110}\bra{001}\right\rangle ,
        \end{aligned}
    \end{equation}

    \begin{equation}
        \frac{d}{dt}\left\langle \hat{a}^{\dagger}\hat{a}\right\rangle =-i\epsilon_{m}\left(t\right)\left\langle \hat{a}\right\rangle ^{*}+i\epsilon_{m}^{*}\left(t\right)\left\langle \hat{a}\right\rangle -\kappa\left\langle \hat{a}^{\dagger}\hat{a}\right\rangle ,
    \end{equation}

    \begin{equation}
        \begin{aligned}
            &\frac{d}{dt}\left\langle \hat{a}\right\rangle \\
            =&-i(\omega_{r}-\omega_{m}+2\alpha \left\langle \hat{a}^{\dagger}\hat{a}\right\rangle+2\sum_{s}\chi_{s}\left\langle \left|s\right\rangle \left\langle s\right|\right\rangle )\\
            &\times\left\langle \hat{a}\right\rangle -i\epsilon_{m}\left(t\right)-\frac{\kappa}{2}\left\langle \hat{a}\right\rangle ,
        \end{aligned}
    \end{equation}

    \begin{equation}
        \begin{aligned}
            &\frac{d}{dt}\left\langle \ket{001}\bra{001}\hat{a}\right\rangle \\
            =&-i\left[\Omega^{*}\left(t\right)\left\langle \ket{001}\bra{110}\hat{a}\right\rangle -\Omega\left(t\right)\left\langle \ket{110}\bra{001}\hat{a}\right\rangle \right]\\
            -&(\gamma_{001\to000}+\frac{\kappa}{2})\left\langle \ket{001} \bra{001}\hat{a}\right\rangle\\
            +&\gamma_{000\to001}\left\langle \ket{000} \bra{000}\hat{a}\right\rangle\\
            -&i\left(\omega_{r}-\omega_{m}+2\alpha \left\langle\hat{a}^{\dagger}\hat{a}\right\rangle+2\chi_{001}\right)\left\langle \ket{001}\bra{001}\hat{a}\right\rangle\\
            -&i\epsilon_{m}\left(t\right)\left\langle \ket{001} \bra{001}\right\rangle ,
        \end{aligned}
    \end{equation}

    \begin{equation}
        \begin{aligned}
            &\frac{d}{dt}\left\langle \ket{110} \bra{110}\hat{a}\right\rangle\\ =&i\left[\Omega^{*}\left(t\right)\left\langle \ket{001} \bra{110}\hat{a}\right\rangle -\Omega\left(t\right)\left\langle \ket{110} \bra{001}\hat{a}\right\rangle \right]\\
            -&(\gamma_{110\to100}+\gamma_{110\to010}+\frac{\kappa}{2})\left\langle \ket{110} \bra{110}\hat{a}\right\rangle \\
            +&\gamma_{100\to110}\left\langle \ket{100} \bra{100}\hat{a}\right\rangle\\
            +&\gamma_{010\to110}\left\langle \ket{010} \bra{010}\hat{a}\right\rangle\\
            -&i\left(\omega_{r}-\omega_{m}+2\alpha \left\langle \hat{a}^{\dagger}\hat{a}\right\rangle+2\chi_{110}\right)\left\langle \ket{110} \bra{110}\hat{a}\right\rangle \\
            -&i\epsilon_{m}\left(t\right)\left\langle \ket{110} \bra{110}\right\rangle ,
        \end{aligned}
    \end{equation}

    \begin{equation}
        \begin{aligned}
            &\frac{d}{dt}\left\langle \ket{100} \bra{100}\hat{a}\right\rangle\\ 
            =&-i\left(\omega_{r}-\omega_{m}+2\alpha \left\langle\hat{a}^{\dagger}\hat{a}\right\rangle+2\chi_{100}\right) \times \\
            &\times\left\langle \ket{100} \bra{100}\hat{a}\right\rangle \\
            -&(\gamma_{100\to000}+\gamma_{100\to110}+\frac{\kappa}{2})\left\langle \ket{100} \bra{100}\hat{a}\right\rangle \\
            +&\gamma_{110\to100}\left\langle \ket{110} \bra{110}\hat{a}\right\rangle \\
            +&\gamma_{000\to100}\left\langle \ket{000} \bra{000}\hat{a}\right\rangle \\
            -&i\epsilon_{m}\left(t\right)\left\langle \ket{100} \bra{100}\right\rangle ,
        \end{aligned}
    \end{equation}

    \begin{equation}
        \begin{aligned}
            &\frac{d}{dt}\left\langle \ket{010} \bra{010}\hat{a}\right\rangle\\ 
            =&-i\left(\omega_{r}-\omega_{m}+2\alpha \left\langle \hat{a}^{\dagger}\hat{a}\right\rangle+2\chi_{010}\right) \times \\
            &\times \left\langle \ket{010} \bra{010}\hat{a}\right\rangle \\
            -&(\gamma_{010\to000}+\gamma_{010\to110}+\frac{\kappa}{2})\left\langle \ket{010} \bra{010}\hat{a}\right\rangle \\
            +&\gamma_{110\to010}\left\langle \ket{110} \bra{110}\hat{a}\right\rangle \\
            +&\gamma_{000\to010}\left\langle \ket{000} \bra{000}\hat{a}\right\rangle \\
            -&i\epsilon_{m}\left(t\right)\left\langle \ket{010} \bra{010}\right\rangle ,
        \end{aligned}
    \end{equation}

    \begin{equation}
        \begin{aligned}
            &\frac{d}{dt}\left\langle \ket{000} \bra{000}\hat{a}\right\rangle\\ 
            =&-i\left(\omega_{r}-\omega_{m}+2\alpha \left\langle \hat{a}^{\dagger}\hat{a}\right\rangle
            +2\chi_{000}\right) \times \\
            & \times \left\langle \ket{000} \bra{000}\hat{a}\right\rangle\\
            -&i\epsilon_{m}\left(t\right)\left\langle \ket{000} \bra{000}\right\rangle \\
            +&\gamma_{100\to000}\left\langle \ket{100} \bra{100}\hat{a}\right\rangle\\
            +&\gamma_{010\to000}\left\langle \ket{010} \bra{010}\hat{a}\right\rangle \\
            +&\gamma_{001\to000}\left\langle \ket{001} \bra{001}\hat{a}\right\rangle \\
            -&\left(\gamma_{000\to100}+\gamma_{000\to010}+\gamma_{000\to001}\right) \times \\
            &\times \left\langle \ket{000} \bra{000}\hat{a}\right\rangle \\
            -&\frac{\kappa}{2}\left\langle \ket{000} \bra{000}\hat{a}\right\rangle ,
        \end{aligned}
    \end{equation}

    \begin{equation}
        \begin{aligned}
            &\frac{d}{dt}\left\langle \ket{110} \bra{001}\hat{a}\right\rangle \\  
            =&i[\left(2\left(\chi_{110}-\chi_{001}\right)\left\langle \hat{a}^{\dagger}\hat{a}\right\rangle -\chi_{001}\right)\left\langle \ket{110} \bra{001}\hat{a}\right\rangle \\
            +&\left(\omega_{110}-\omega_{001}-\omega_{p}\right)\left\langle \ket{110} \bra{001}\hat{a}\right\rangle ]\\
            +&i\Omega^{*}\left(t\right)\left[\left\langle \ket{001} \bra{001}\hat{a}\right\rangle -\left\langle \ket{110} \bra{110}\hat{a}\right\rangle \right]\\
            -&[\frac{1}{2}\left(\gamma_{110\to100}+\gamma_{110\to010}+\gamma_{001\to000}\right)\\
            +&\gamma_{\phi}+\frac{\kappa}{2}]\left\langle \ket{110} \bra{001}\hat{a}\right\rangle \\
            -&i\left(\omega_{r}-\omega_{m}+2\alpha \left\langle \hat{a}^{\dagger}\hat{a}\right\rangle\right)\left\langle \ket{110} \bra{001}\hat{a}\right\rangle \\
            -&i\epsilon_{m}\left(t\right)\left\langle \ket{110} \bra{001}\right\rangle ,
        \end{aligned}
    \end{equation}

    \begin{equation}
        \begin{aligned}
            &\frac{d}{dt}\left\langle \ket{001} \bra{110}\hat{a}\right\rangle \\  
            =&i[\left(2\left(\chi_{001}-\chi_{110}\right)\left\langle \hat{a}^{\dagger}\hat{a}\right\rangle -\chi_{110}\right)\left\langle \ket{001} \bra{110}\hat{a}\right\rangle \\
            -&\left(\omega_{110}-\omega_{001}-\omega_{p}\right)\left\langle \ket{001} \bra{110}\hat{a}\right\rangle ]\\
            -&i\Omega\left(t\right)\left[\left\langle \ket{001} \bra{001}\hat{a}\right\rangle -\left\langle \ket{110} \bra{110}\hat{a}\right\rangle \right]\\
            -&[\frac{1}{2}\left(\gamma_{110\to100}+\gamma_{110\to010}+\gamma_{001\to000}\right)\\
            +&\gamma_{\phi}+\frac{\kappa}{2}]\left\langle \ket{001} \bra{110}\hat{a}\right\rangle \\
            -& i\left(\omega_{r}-\omega_{m}+2\alpha \left\langle \hat{a}^{\dagger}\hat{a}\right\rangle\right)\left\langle \ket{001} \bra{110}\hat{a}\right\rangle \\
            -&i\epsilon_{m}\left(t\right)\left\langle \ket{001} \bra{110}\right\rangle .
        \end{aligned}
    \end{equation}

\label{eq:cavity_bloch}
\end{subequations}
Here, dissipation in the system is described by the total decay rate of the resonator $\kappa=\kappa_{\mathrm{int}}+\kappa_{\mathrm{ext}}$, the dephasing rate $\gamma_{\phi}$ between $\ket{110}$ and $\ket{001}$, and the qubit state decay rates $\gamma_{i\to f}$ with the following pairs of initial and final states $(\ket{i}, \ket{f})$: $(\ket{110}, \ket{100})$, $(\ket{110}, \ket{010})$, $(\ket{100}, \ket{000})$, $(\ket{010}, \ket{000})$, and $(\ket{001}, \ket{000})$ and the transition rates with $\ket{i}$ and $\ket{f}$ interchanged.

We first determine the shape of the pulse $\epsilon_m(t)$ for each resonator by measuring and fitting to the time-dependent resonator response at different readout frequencies $\omega_m$ in the qubit ground state $\ket{000}$, using the values of $\omega_r$, $\kappa_{\mathrm{int}}$ and $\kappa_{\mathrm{ext}}$ obtained from the VNA measurements in Table~\ref{tab:Resonators}. Then, assuming the qubit state before the $\pi$ pulse on qubit 3 is an incoherent mixture of $\ket{000}$, $\ket{001}$, $\ket{100}$ and $\ket{010}$, where all single-photon populations are thermal, we fit the time-dependent resonator response for all three resonators, different readout frequencies $\omega_m$ and different evolution times $t_{\mathrm{evolve}}$. The results are shown in Fig.~\ref{fig:Measurements}.

\section{\label{app:string}Charge and Parity Symmetries of the gauge invariant model and real-time evolution of a string}
Apart from local gauge invariance, $[\hat{\mathcal{H}}_m, \hat{G}_n ]=0$ for all sites $n$, where Eq.~\eqref{eq:H} gives the Hamiltonian of the model and Eq.~\eqref{eq:Gsymm} the local gauge generators respectively, two extra symmetries characterize the model: parity and charge conjugation.

The first one, parity, is defined on the operators by
\begin{equation}
\begin{split}
\hat{\sigma}^z_n  \mapsto \hat{\sigma}^z_{-n};& ~~ \hat{\sigma}_{n}^{+} \mapsto \hat{\sigma}_{-n}^{+}\\
\hat{\tau}^z_{n,n+1}  \mapsto -\hat{\tau}^z_{-n-1,-n};&  ~~ \hat{\tau}^+_{n,n+1} \mapsto \hat{\tau}^-_{-n-1,-n}
\end{split}
\end{equation}
The second one, charge conjugation, is defined on the operators by 
\begin{equation}
\begin{split}
\hat{\sigma}^z_n  \mapsto -\hat{\sigma}^z_{n+1};& ~~ \hat{\sigma}_{n}^{+} \mapsto \hat{\sigma}_{n+1}^{-}\\
\hat{\tau}^z_{n-1,n}  \mapsto -\hat{\tau}^z_{n,n+1};&  ~~ \hat{\tau}^+_{n-1,n} \mapsto \hat{\tau}^-_{n,-n+1}.
\end{split}
\end{equation}
\begin{figure}[htbp]
    \centering
    \includegraphics[scale = 0.5]{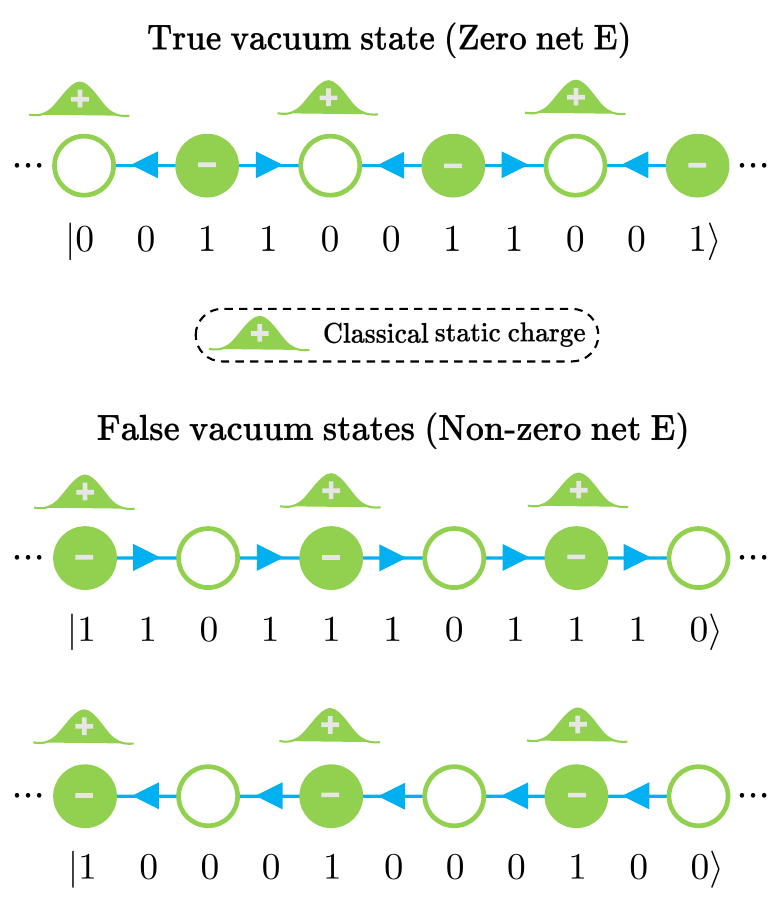}
    \caption{ Representation of the three central states for the false vacuum decay of the system in the strong coupling limit $|\mu|\gg|J|$. The total net electric flux $\hat{E}$ acts as the order parameter of two phases, where the parity and charge conjugation symmetries get broken. The upper state is the true vacuum of the gauge-invariant system when $\mu/J \gg 1$, having the staggered fermions the rightmost site occupied and preserving the symmetries. Here the classical and quantum charges act as sources and sinks of the electric field. The bottom states, where the fermions have the leftmost site occupied, are the false vacua of the system at $\mu/J \gg 1$. They form the two-fold degenerate phase where the parity and charge conjugation symmetries are broken, having a non-zero net electric flux in two possible directions. Here the charges get canceled out or the site is directly empty, so there are no sources/sinks to change the course of the electric field. Notice the states have been represented as an example in a 6-site 1D lattice. }  
    \label{fig:Cartoon_states}
\end{figure}
%

%These definitions of the symmetry transformations can be applied to the three states that minimize the energy, which are depicted in Fig. \ref{fig:Cartoon_states} and defined in the strong coupling limit, i.e. when $|\mu|\gg|J|$. 

These definitions of the symmetry transformations can be applied to the three states that minimize the energy in the strong coupling limit (i.e. when $|\mu / J |\gg1$), depicted in Fig. \ref{fig:Cartoon_states}. 

The first state we characterize is a charge and parity-preserving state: when $\mu / J \gg 1$, the odd matter sites are empty and the even matter sites are occupied. In addition to this and due to Gauss' law, the electric flux at every link points towards the empty sites and leaves from the occupied sites, giving a zero total electric flux. It is easy to realize that this state preserves the charge and parity symmetries.

The second pair of states are charge and parity-breaking states: they are the ground states when $-\mu / J \gg 1$, so that the odd matter sites are occupied and the even matter sites are empty. In addition to this and due to Gauss' law, the electric flux at every link points in the same direction giving rise to two possible states: if the electric flux is pointing to the right or the left along the string. This pair of states are charge and parity conjugates, i.e. one is transformed into the other by the action of any symmetry operations.

Having these states in mind, the process of false vacuum decay can be tracked by a real-time evolution of our system. The system progressively transitions to the true vacuum, beginning with one of the false vacua configurations and undergoing gauge invariant time evolution. This can be seen for $\mu / J = 0$ in Fig.~\ref{fig:NandE_oddeven_other} of the main text by the study of the particle number $\hat{N}_n = ( \mathbb{1} - \hat{\sigma}^z_n )/2 $ per (matter) site and the electric field $\hat{E}_{n} = -\hat{\tau}^z_{n,n+1} $ per (gauge) link. 

\begin{figure}[htbp]
    \centering
    \includegraphics[scale=0.479]{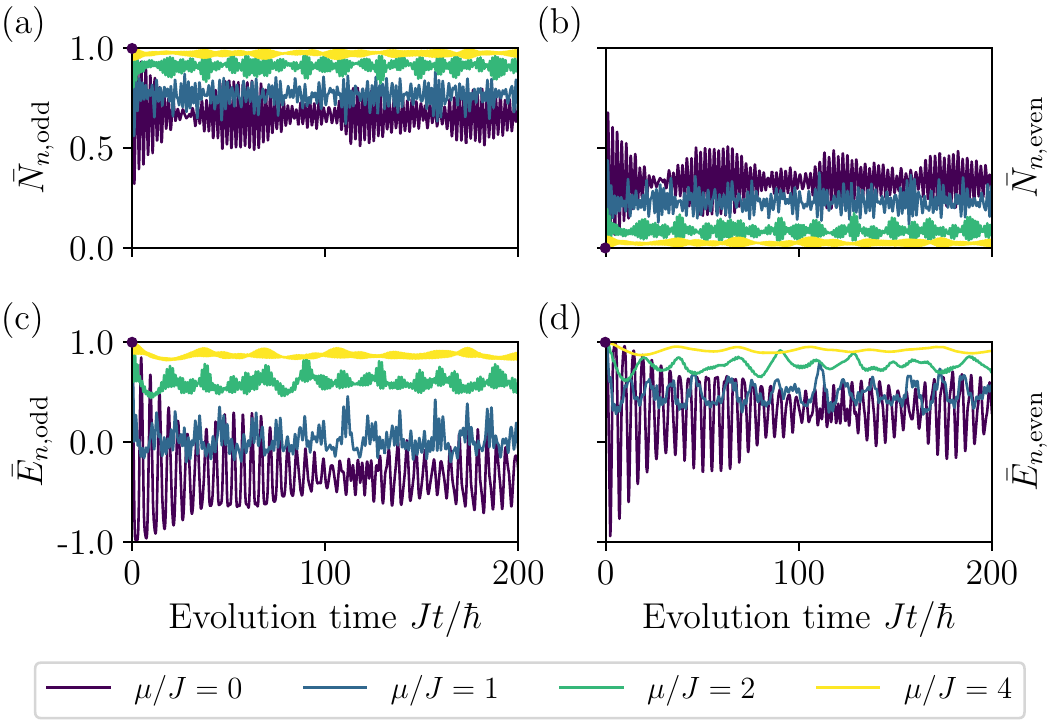}
    \caption{Effects of an increasing site energy $\mu$ on the system's time evolution from a false vacuum configuration. (a)/(b) Particle number $\hat{N}_n$ per odd/even matter sites of the 1D lattice. (c)/(d) Electric field $\hat{E}_n$ per odd/even links of the 1D lattice. A staggered particle number configuration and non-zero electric flux (same direction in odd and even links) are maintained as we increase the site energy $\mu$, leaving the system in the symmetry-breaking configuration. These can be seen with the values of $\bar{N}_{n,\rm odd} \rightarrow 1$, $\bar{N}_{n,\rm even} \rightarrow 0$, $\bar{E}_{n,\rm odd} \rightarrow 1 $ and $ \bar{E}_{n,\rm even} \rightarrow 1$, matching one of the false vacuum configurations in Fig. \ref{fig:Cartoon_states}. The most unstable evolution is shown for site energy $\mu/J = 0$, where the starting values are marked with a purple circle. The averaged magnitudes have been calculated with the values of the three central odd and even sites/links acting as the bulk, leaving the rest as the environment. The simulation is done with a 12-site lattice for $Jt/\hbar = 200$.} 
    \label{fig:Averages_increasing_mu_false}
\end{figure}

\begin{figure}[htbp]
    \centering
    \includegraphics[scale=0.479]{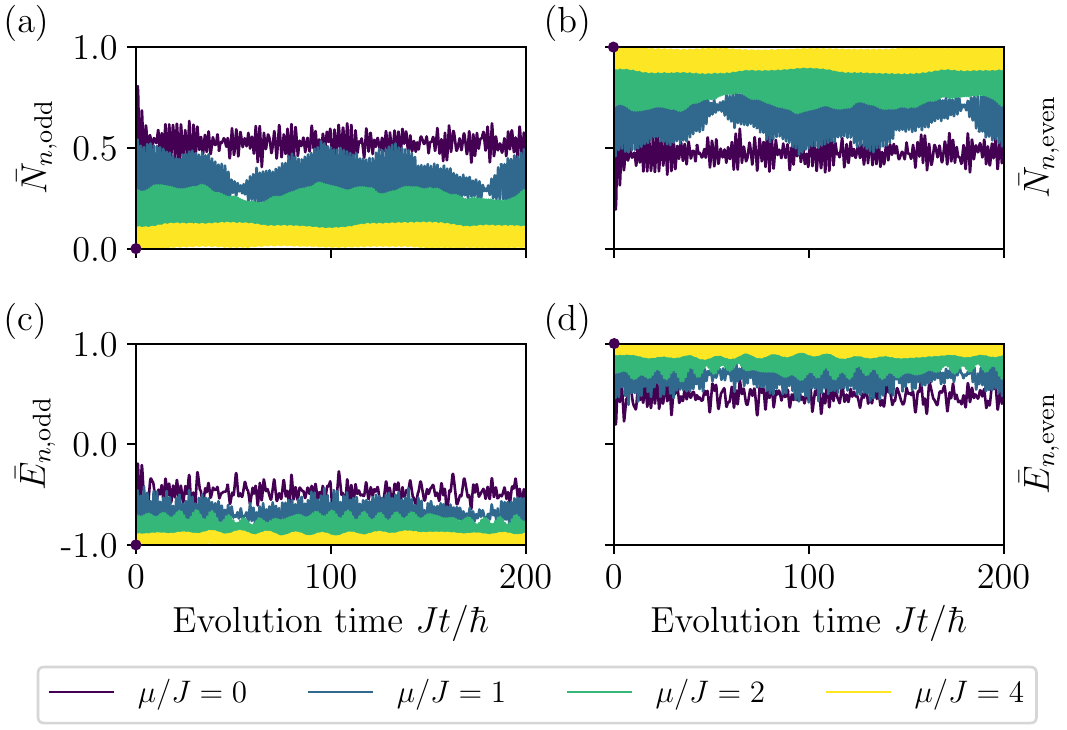}
    \caption{Effects of an increasing site energy $\mu$ on the system's time evolution from the true vacuum configuration. (a)/(b) Particle number $\hat{N}_n$ per odd/even matter sites of the 1D lattice. (c)/(d) Electric field $\hat{E}_n$ per odd/even links of the 1D lattice. A staggered particle number configuration and zero net electric flux (opposite direction in odd and even links) are maintained as we increase the site energy $\mu$, leaving the system in the symmetry-preserving configuration i.e. the true vacuum. These can be seen with the values of $\bar{N}_{n,\rm odd} \rightarrow 0$, $\bar{N}_{n,\rm even} \rightarrow 1$, $\bar{E}_{n,\rm odd} \rightarrow -1 $ and $ \bar{E}_{n,\rm even} \rightarrow 1$, matching the true vacuum configuration in Fig. \ref{fig:Cartoon_states}. The most different evolution is shown for site energy $\mu/J = 0$, where the starting values are marked with a purple circle. The averaged magnitudes have been calculated with the values of the three central odd and even sites/links acting as the bulk, leaving the rest as the environment. The simulation is done with a 12-site lattice for $Jt/\hbar = 200$.} 
    \label{fig:Average_increasing_mu_true}
\end{figure}

However, suppose we evolve the initial configuration with a different increasing value of the site energy $\mu / J$. In that case, we observe that the state tends to remain in the configuration we started with, whether the symmetry-breaking or the symmetry-preserving one. These can be observed in Figs.~\ref{fig:Averages_increasing_mu_false} and \ref{fig:Average_increasing_mu_true} respectively.

%

%\newpage
\clearpage

\bibliography{bibl, Notes}

\end{document}